# Visualizing heavy fermions emerging in a quantum critical Kondo lattice

Pegor Aynajian[1]*, Eduardo H. da Silva Neto[1]*, András Gyenis[1], Ryan E. Baumbach[2], J. D. Thompson[2], Zachary Fisk[3], Eric D. Bauer[2], and Ali Yazdani[1]§

[1]Joseph Henry Laboratories and Department of Physics, Princeton University, Princeton, NJ 08544 USA
[2]Los Alamos National Laboratory, Los Alamos, New Mexico 87545, USA
[3]University of California, Irvine, California 92697, USA

§To whom correspondence should be addressed. Email: yazdani@princeton.edu

*These authors contributed equally to this work.

**In solids containing elements with *f*-orbitals, the interaction between *f*-electron spins and those of itinerant electrons leads to the development of low-energy fermionic excitations with a heavy effective mass. These excitations are fundamental to the appearance of unconventional superconductivity and non-Fermi liquid behavior observed in actinide- and lanthanide-based compounds. We use spectroscopic mapping with the scanning tunneling microscope to detect the emergence of heavy excitations with lowering of temperature in a prototypical family of Ce-based heavy fermion compounds. We demonstrate the sensitivity of the tunneling process to the composite nature of these heavy quasiparticles, which arises from quantum entanglement of itinerant conduction and *f*-electrons. Scattering and interference of the composite quasiparticles is used to resolve their energy-momentum structure and to extract their mass enhancement, which develops with decreasing temperature. Remarkably, the lifetime of the emergent heavy quasiparticles reveals signatures of enhanced scattering and their spectral lineshape shows evidence of energy-temperature scaling. These findings demonstrate that proximity to a quantum critical point results in critical damping of the emergent heavy excitation of our Kondo lattice system.**

A local magnetic moment occurs when a strongly interacting quantum state, such as an atomic *d*- or *f*-orbital, cannot be doubly occupied due to strong on-site Coulomb repulsion[1]. In the presence of a dilute concentration of such magnetic moments in a metal, spin-flip scattering of conduction electrons from these local moments results in their collective magnetic screening below a characteristic temperature called the Kondo temperature $T_K$ (Fig. 1a)[2]. In materials where local moments are arranged in a dense periodic array, the so-called Kondo lattice, the deconfinement of localized orbitals through their hybridization with the conduction electrons results in composite low-energy excitations with a heavy effective mass (Fig. 1b). Tuning the hybridization between *f*-orbitals and itinerant electrons can destabilize the heavy Fermi liquid state towards an antiferromagnetically ordered ground state[3-8]. In proximity to such a quantum phase transition, between itinerancy and localization of *f*-electrons, many heavy fermion systems exhibit unconventional superconductivity at low temperatures (Fig. 1c)[9].

Thermodynamic and transport studies have long provided evidence for heavy quasiparticles, their unconventional superconductivity and non-Fermi liquid behavior in a variety of material systems[9-14]. However, the emergence of a coherent band of heavy quasiparticles near the Fermi energy in a Kondo lattice system is still not well understood[14-17]. Part of the challenge has been the inability of spectroscopic measurements to probe the development of heavy quasiparticles with lowering of temperature and to characterize their properties with high-energy resolution. Such precise measurements of heavy fermion formation are not only required for understanding the nature of these electronic excitations close to quantum phase transitions[18] but are also critical to identifying the source of unconventional superconductivity near such transitions.

*Composite heavy fermion excitations*

The emergence of composite heavy fermions in a Kondo lattice can be considered as a result of the hybridization of two electronic bands: one dispersing band due to conduction electrons and one weakly dispersing band originating from localized *f*-electrons (dashed lines in Fig. 1d). This hybridization generates low-energy quasiparticles that are a mixture of conduction- and *f*-electrons with a modified band structure characterized by the so-called direct ($2v$) and indirect ($\Delta_h$) hybridization gaps, as shown in Fig. 1d[17,19]. Various theoretical approaches, including several numerical studies, remarkably reproduce the generic composite band structure in Fig. 1d[20-24]. Recent theoretical modeling has also shown that tunneling spectroscopy can be a powerful probe of this composite nature of heavy fermions[25-28]. Depending on the relative tunneling amplitudes to the light conduction ($t_c$) or to the heavy *f*-like ($t_f$) components of the composite quasiparticles, and due to their interference, tunneling spectroscopy can be sensitive to different features of the hybridized band structure. Figures 1d-f show examples of model calculations (see Supplementary section I) illustrating the sensitivity to predominantly tunneling to the light (Fig. 1e) or heavy (Fig. 1f) electronic states.

Recent advances in the application of STM to heavy fermions are providing a new approach to examining the correlated electrons in these systems with high energy and spatial resolutions. STM and point-contact experiments on heavy fermion compounds have shown evidence for hybridization of the conduction electrons with the *f*-orbitals and have been used to probe the so-called hidden order phase transition involving heavy $f$-electrons in $URu_2Si_2$[29-32]. Sudden onset of the hidden order phase appears to give rise to strong modification of the band structure in $URu_2Si_2$ as detected by STM measurements[30,31]. However, these changes are correlated with the phase transition into the hidden order at 17.5 K rather than the generic physics of heavy Fermi liquids that should appear at higher temperatures and evolve smoothly with lowering of temperature. Direct experimental observation of the gradual formation of heavy quasiparticles with temperature and evidence of their composite nature, which is ubiquitous to all heavy fermions, as well as examination of their properties in proximity to quantum critical points (QCPs), have remained out of the reach of STM and other spectroscopic measurements.

### *$Ce_1M_1In_5$ as a model heavy fermion system*

To provide a controlled study of the emergence of heavy fermion excitations within a Kondo lattice system that can be tuned close to a QCP, we carried out studies on the $Ce_1M_1In_5$ (with M = Co, Rh) material system. These so-called 115 compounds offer the possibility to tune the interaction between the Ce's $f$-orbitals and the itinerant *spd* conduction electrons using isovalent substitutions at the transition metal site within the same tetragonal crystal structure. Consequently, the ground state of this system can be tuned (in stoichiometric compounds) between antiferromagnetism, as in $CeRhIn_5$ ($T_N$ = 3.5 K), to superconductivity, as observed in $CeCoIn_5$ ($T_c$ = 2.3 K) and $CeIrIn_5$ ($T_c$ = 0.4 K)[9]. Previous studies indicate that $CeCoIn_5$ is very close to a QCP[33-36], while $CeRhIn_5$ can be tuned close to this transition with application of pressure[7,37]. These experiments confirm that superconductivity in the 115 system emerges at low temperatures close to a QCP from the development of heavy low-energy excitations at high temperature[7,9,36]. More specifically, transport studies show a drop in the electrical resistivity in CeCoIn$_5$ around 50 K, which has been interpreted as evidence for the development of a coherent heavy quasiparticle band, followed by a linear resistivity at lower temperature (above $T_c$)[38] — a behavior that has been associated with the proximity to the QCP. Quantum oscillations and thermodynamic measurements find a heavy effective mass (10-50 $m_0$, bare electron mass) for CeCoIn$_5$, while in the same temperature range the *f*-electrons in CeRhIn$_5$ are effectively decoupled from the conduction electrons[39,40].

Figure 2 shows STM images of a single crystal of CeCoIn$_5$ that has been cleaved *in situ* in our variable temperature ultra-high vacuum STM. In this family of compounds the cleaving process results in exposing multiple surfaces terminated with different chemical compositions. The crystal symmetry necessarily requires multiple surfaces for cleaved samples, as no two equivalent consecutive layers occur within the unit cell. Therefore breaking of any single chemical bond will result in different layer terminations on the two sides of the cleaved sample. Experiments on multiple cleaved samples have revealed three different surfaces, two of which are atomically ordered (termed surfaces A and B in Figs. 2a,b) with a periodicity of ~ 4.6 Å corresponding to the lattice constant of the bulk crystal structure, while the third surface (termed surface C , Fig. 2b) is reconstructed. Comparison of the relative heights of the sub-unit cell steps between the different layers (Figs. 2c,d) to the crystal structure determined from scattering experiments[41] suggests that the exposed surfaces A, B, and C correspond to the Ce-In, Co, and In$_2$ layers, respectively. Experiments on CeCo(In$_{0.9985}$Hg$_{0.0015}$)$_5$ and CeRhIn$_5$ reveal similar results, where cleaving exposes the corresponding multiple layers in those compounds (see Supplementary section II). Hg defects in CeCoIn$_5$ at this concentration have negligible influence on its thermodynamic and transport properties and are introduced for scattering experiments described below[42].

*Signatures of hybridization and composite excitations*

Spectroscopic measurements of CeCoIn$_5$ show the sensitivity of the tunneling process to the composite nature of the hybridized heavy fermion states. As shown in Fig. 3a, tunneling spectra on surface A (identified as the Ce-In layer) of CeCo(In$_{0.9985}$Hg$_{0.0015}$)$_5$ and CeCoIn$_5$ (shown in Supplementary section III) show that upon cooling the sample dramatic changes develop in the spectra in an asymmetric fashion about the Fermi energy. The redistribution of the spectra observed on this surface is consistent with a tunneling process that is dominated by coupling to the light conduction electrons and displays signatures of the direct hybridization gap (2*v*) experienced by this component of the heavy fermion excitations (e.g. see Figs. 1d,e). In contrast to these observations, similar measurements on the corresponding surface of CeRhIn$_5$ show spectra that are featureless in the same temperature range (Fig. 3a, dashed line) and are consistent with the more localized nature of the Ce *f*-orbitals in CeRhIn$_5$ as compared to CeCoIn$_5$. The hybridization gap structure in CeCoIn$_5$ is also centered above the chemical potential (8 meV, see Fig. 3a), which makes it difficult for angle-resolved photoemission experiments[43-45], the typical technique to probe electronic band structure in solids, to access.

The composite nature of the heavy fermion excitations manifests itself by displaying different spectroscopic characteristics for tunneling into the different atomic layers. Figure 3b shows spectra measured on surface B (identified as Co) of CeCo(In$_{0.9985}$Hg$_{0.0015}$)$_5$ that looks remarkably different than those measured on surface A (Fig. 3a). In the same temperature range where spectra on surface A (Fig. 3a) develop a depletion of spectral weight near the Fermi energy, surface B shows a sharp enhancement of spectral weight within the same energy window (Fig. 3b). With further lowering of temperature, the enhanced tunneling on surface B evolves into a double-peak structure. As a control experiment, measurements on the corresponding surface in CeRhIn$_5$, once again, display no sharp features in the same temperature and energy windows (Fig. 3b, dashed line). The spectroscopic features of CeCo(In$_{0.9985}$Hg$_{0.0015}$)$_5$'s surface B display the characteristic signatures of dominant tunneling to the *f*-component of the heavy quasiparticles, which reside near the Fermi energy and are expected to display the indirect hybridization gap ($\Delta_h$) (see Fig. 1d,f).

A model calculation for tunneling to composite heavy excitations can reproduce our spectroscopic measurements on the two different atomically ordered surfaces of CeCo(In$_{0.9985}$Hg$_{0.0015}$)$_5$. Following recent theoretical efforts[26,27], we compute spectroscopic properties of a model band structure in which a single hole-like itinerant band of *spd*-like electrons hybridizes with a narrow band of *f*-like electrons (see Supplementary section I for details of the model). The results of our calculations (Fig. 3c,d) are sensitive to the ratio of tunneling ($t_f/t_c$) into the heavy *f*-states versus the light conduction band — a behavior that explains the differences between the tunneling processes on the different cleaved surfaces (Fig. 3a,b). While naively one would expect that tunneling to the heavy excitations would be more pronounced on the Ce-In layer, recent first principles calculations show that the amplitude of the hybridization of the *f*-states with the out-of-plane spd-electrons can be remarkably larger than with those in-plane[21].

*Visualizing quasiparticle mass enhancement*

To directly probe the energy-momentum structure of heavy quasiparticles in the 115 material systems, we have carried out spectroscopic mapping with the STM that enables us to visualize scattering and interference of these quasiparticle excitations from impurities or structural defects. Elastic scattering of quasiparticles from these imperfections gives rise to standing waves in the conductance maps at wavelengths corresponding to *2π/q* where *q = k$_f$ - k$_i$* is the momentum transfer between initial (*k$_i$*) and final (*k$_f$*) states at the same energy. We expect that *q*'s with the strongest intensity connect regions of high density of states on the contours of constant energy and hence provide energy-momentum information of the quasiparticle excitations. We characterize the scattering *q*'s using discrete Fourier transforms (DFTs) of STM conductance maps measured at different energies. The presence of Hg substitutions in CeCo(In$_{0.9985}$Hg$_{0.0015}$)$_5$ provides a sufficient number of scattering centers to enhance signal to noise ratio for such quasiparticle interference (QPI) measurements.

Figure 4a shows examples of energy-resolved STM conductance maps on surface A of CeCo(In$_{0.9985}$Hg$_{0.0015}$)$_5$ measured at 20 K displaying signatures of scattering and interference of quasiparticles from defects and step edges. These conductance maps show clear changes of the wavelength of the modulations as a function of energy. Perhaps the most noticeable are the changes around each random defect (see Supplementary section II for the corresponding STM image showing the location of the Hg defects). Figure 4b shows

DFTs of such maps that display sharp non-dispersive Bragg peaks (at the corners, (±2π/a,0), (0,±2π/a)) corresponding to the atomic lattice, as well as other features (concentric square-like shapes) that rapidly disperse with energy, collapse (Fig. 4b; 0 meV), and disappear (Fig. 4b; 9 meV) near the Fermi energy. We have carried out such measurements both at low temperatures (20 K, Fig. 4b) where the spectrum shows signatures of hybridization between conduction electrons and *f*-orbitals, as well as at high temperatures (70 K, Fig. 4c) where such features are considerably weakened (e.g. Fig. 4c; 2 meV, 10 meV). As a control experiment, we have also carried out the same measurements on the corresponding surface of $CeRhIn_5$ (Fig. 4d), for which signatures of heavy electron behavior are absent (e.g. Fig. 3a) in the same temperature window (20 K). While understanding details of the QPI in Fig. 4 requires detailed modeling of the band structure of the 115 compounds, the square-like patterns observed in the data correspond to scattering wavevectors that can be identified from the calculated LDA band structure[46] (see Supplementary section V).

We find that analyzing the features of the energy-resolved DFT maps provide direct evidence for mass enhancement of quasiparticles, in unison with related signatures in the tunneling spectra. Figures 5a,b show line cuts of the DFT maps plotted along two high symmetry directions as a function of energy for $CeCo(In_{0.9985}Hg_{0.0015})_5$ at 20 K and in Fig. 5c we show their corresponding spatially averaged spectrum. The square-like regions of enhanced quasiparticle scattering in Fig. 4b appear in the line cuts of Figs. 5a,b as energy-dependent bands of scattering, which become strongly energy dependent near the Fermi energy. Clearly the scattering of the quasiparticle excitations in the energy window near the direct hybridization gap have flatter energy-momentum structure as compared to those at energies away from the gap. This is the direct signature of the quasiparticles acquiring heavy effective mass at low energies near the Fermi energy. Detailed analysis of one of the QPI bands estimates the mass enhancement near the Fermi energy to be about $30m_0$ (Fig. 5a inset), a value which is close to that seen in quantum oscillation studies of $CeCoIn_5$[39,40]. Our model calculation, which describes the spectroscopic lineshapes on the different surfaces, can also be extended to reproduce the signatures of mass enhancement in the QPI data (see Supplementary section I).

Contrasting low temperature QPI patterns on $CeCo(In_{0.9985}Hg_{0.0015})_5$ to measurements on the same compound at high temperatures (70 K, Figs. 5d,e), where the hybridization gap is weak (Fig. 5f), or to measurements on $CeRhIn_5$ (20 K, Fig. 5g,h), where signatures of a hybridization gap are absent in the tunneling spectra (Fig. 5i), confirms that the development of this gap results in apparent splitting of the bands which are responsible for both the scattering and the heavy effective mass in the QPI measurements. Furthermore, these measurements show that the underlying band structure responsible for the scattering wavevectors away from the Fermi energy is relatively similar between $CeCo(In_{0.9985}Hg_{0.0015})_5$ and $CeRhIn_5$. Only when *f*-electrons of the Kondo lattice begin to strongly hybridize with conduction electrons and modify the band structure within a relatively narrow energy window (30 meV), we see signatures of heavy fermion excitations in QPI measurements, signaling a transition from small to large Fermi surface (see Supplementary section V).

*Quasiparticle lifetime & signatures of quantum criticality*

The ability to tunnel through the *f*-component of the heavy quasiparticles on CeCo(In$_{0.9985}$Hg$_{0.0015}$)$_5$'s surface B provides an opportunity to probe the lifetime of the heavy quasiparticles as a function of temperature in a system that is close to a QCP. The narrow dispersion of the *f*-band results in a direct connection between the experimentally measured width of the peak in the density of states near the Fermi energy (Fig. 3b) and lifetime of the heavy quasiparticles. Analysis of this width measured at different temperatures is displayed in Figure 6a (see Supplementary section VI) and shows a strong temperature dependence with a finite intercept (~ 3.5meV) in the limit of zero temperature. The finite width at zero temperature can be understood as a consequence of a small but finite dispersion of the *f*-band as well as a finite probability of tunneling into the *spd* electrons (see SI section-I). However, the large linear slope in Figure 6a (larger than 3/2 $k_BT$) indicates that the *f*-electron's lifetime, as opposed to thermal broadening, is strongly influencing the spectra and its temperature dependence. Consistent with this observation, we also find that to capture the temperature evolution of the spectra in Fig. 3b, we have to use rather large values of scattering rate (inverse lifetime) of the *f*-component of the heavy quasiparticles $\gamma_f = \hbar/\tau_f$ in our model calculations (Fig. 3d).

A linear scattering rate (or inverse lifetime) for the heavy quasiparticles is consistent with the expectation that CeCoIn$_5$ is close to a QCP, since for systems tuned close to such transitions, temperature is the only relevant energy scale available to determine the quasiparticle lifetime, resulting in $\hbar/\tau_f \propto k_BT$ [18,47]. Yet, a more precise signature of a QCP would be the observation of energy-temperature scaling of experimental quantities near such transitions. In fact, recent theoretical work suggests that the instability of the Fermi surface near a QCP should result in scaling properties of the single-particle excitation that can be directly probed in measurements of the tunneling density of states[48]. To test this hypothesis, we examine the lineshape of the tunneling spectra on surface B near the chemical potential at different temperatures and attempt to scale the data (Fig. 3b) by plotting $(dI/dV)_S*(k_BT^{\alpha})$ as a function of $E/(k_BT)^{\beta}$ where $(dI/dV)_S$ is the background subtracted spectra of Fig. 3b (see supplementary section VI) and $E$ is the energy of the tunneling quasiparticles. We find that using the exponents $\alpha$ = 0.53 and $\beta$ = 1 results in a collapse of the data at different temperatures on a single curve in the low bias region (Fig. 6b and see supplementary section VII). Although an understanding of the value of the exponent $\alpha$ is currently lacking, the linear power $\beta$ confirms our hypothesis of energy-temperature scaling associated with proximity to a QCP. These results indicate that the heavy quasiparticles in CeCoIn$_5$ are damped because of critical fluctuations rather than the typical scattering that is expected in a Fermi liquid ($T^2$ dependence). Similar energy-temperature scalings, with anomalous exponents ($\alpha$), have been previously observed in the dynamical spin susceptibility of other heavy fermion systems near QCPs[3,5,49,50]. However, here we show for the first time that the signatures of scaling and critical phenomena appear in the spectroscopic properties of the quasiparticle excitations.

## *Conclusion and outlook*

The experimental results and the model calculations presented here provide a comprehensive picture of how the heavy fermion excitations in the 115 Ce-based Kondo lattice systems emerge with lowering of temperature or as a result of chemical tuning of the interaction between the Ce *f*-electrons and the conduction electrons. The changes in the scattering properties of the quasiparticles directly signal the flattening of their energy-momentum structure and the emergence of heavy quasiparticles near the Fermi energy. Such changes are also consistent with the predicted evolution from small to large Fermi surface as the localized *f*-electrons hybridize with the conduction electrons. The sensitivity of the tunneling to the surface termination and the successful modeling of these data provide direct spectroscopic evidence of the composite nature of heavy fermions and offer a unique method to disentangle their components.

Our experiments also demonstrate that the emergent heavy quasiparticles in our system are strongly scattered and show signatures of scaling associated with critical damping of excitations in proximity to a QCP. Like many other heavy fermion systems, thermodynamic and transport studies of the 115 systems have shown evidence of quantum criticality, but never before such signatures have been isolated in an electron spectroscopy measurement as described here. Such spectroscopic signatures are a direct evidence for the breakdown of coherent fermionic excitations approaching a QCP. Future extension of our measurements to lower temperatures can probe the interplay between quantum fluctuations and the appearance of superconductivity, an issue which continues to be one of the most debated in condensed matter physics.

***Methods Summary***

The single crystals of $CeCoIn_5$, $CeCo(In_{0.9985}Hg_{0.0015})_5$, and $CeRhIn_5$ used for this study were grown from excess indium at Los Alamos National Laboratory. Small, flat crystals were oriented along the crystallographic axes and cut into sizes suitable for STM measurements (~2 x 2 x 0.2 mm$^3$). The samples were cleaved on a surface perpendicular to the c-axis in ultra-high vacuum (UHV) and transferred in situ to the microscope head. Differential conductance (*dI/dV*) measurements were performed using standard lock-in techniques. Approximately ten different samples of CeCoIn$_5$, CeCo(In$_{0.9985}$Hg$_{0.0015}$)$_5$, and CeRhIn$_5$ were successfully cleaved and studied and the spectroscopic data collected were reproducible on the corresponding identical exposed surfaces of the different samples. Spectra measured at different locations on each surface showed negligible variations. The spectra presented in the main manuscript are averaged over approximately 200 individual spectra measured over an area of at least 100 Å x 100 Å. The spectroscopic lineshapes showed negligible variations as the tip height was varied (variation of the tunneling current by two orders of magnitude).

# References

1   Anderson, P. W. Localized Magnetic States in Metals. *Phys. Rev.* **124**, 41-53 (1961).
2   SPECIAL TOPICS: Kondo Effect - 40 Years after the discovery. *J. Phys. Soc. Jpn.* **74**, 1-238 (2005).
3   Schroder, A. *et al.* Onset of antiferromagnetism in heavy-fermion metals. *Nature* **407**, 351-355 (2000).
4   Coleman, P., Pépin, C., Si, Q. & Ramazashvili, R. How do Fermi liquids get heavy and die? *J. Phys. Cond. Matter* **13**, R723-R738 (2001).
5   Si, Q., Rabello, S., Ingersent, K. & Smith, J. L. Locally critical quantum phase transitions in strongly correlated metals. *Nature* **413**, 804-808 (2001).
6   Senthil, T., Sachdev, S. & Vojta, M. Fractionalized Fermi Liquids. *Phys. Rev. Lett.* **90**, 216403 (2003).
7   Park, T. *et al.* Hidden magnetism and quantum criticality in the heavy fermion superconductor $CeRhIn_5$. *Nature* **440**, 65-68 (2006).
8   Gegenwart, P., Si, Q. & Steglich, F. Quantum criticality in heavy-fermion metals. *Nat Phys* **4**, 186-197 (2008).
9   Pfleiderer, C. Superconducting phases of f-electron compounds. *Rev. Mod. Phys.* **81**, 1551-1624 (2009).
10  Palstra, T. T. M. *et al.* Superconducting and Magnetic Transitions in the Heavy-Fermion System $URu_2Si_2$. *Phys. Rev. Lett.* **55**, 2727-2730 (1985).
11  Stewart, G. Heavy-fermion systems. *Rev. Mod. Phys.* **56**, 755-787 (1984).
12  Fisk, Z., Sarrao, J. L., Smith, J. L. & Thompson, J. D. The physics and chemistry of heavy fermions. *Proc. Natl. Acad. Sci. U.S.A.* **92**, 6663-6667 (1995).
13  Steglich, F. *et al.* Classification of strongly correlated f-electron systems. *J. Low Temp. Phys.* **99**, 267-281 (1995).
14  Yang, Y.-f., Fisk, Z., Lee, H.-O., Thompson, J. D. & Pines, D. Scaling the Kondo lattice. *Nature* **454**, 611-613 (2008).
15  Si, Q. & Steglich, F. Heavy Fermions and Quantum Phase Transitions. *Science* **329**, 1161-1166 (2010).
16  Anderson, P. W. Fermi Sea of Heavy Electrons (a Kondo Lattice) is Never a Fermi Liquid. *Phys. Rev. Lett.* **104**, 176403 (2010).
17  Coleman, P. in *Handbook of Magnetism and Advanced Magnetic Materials* Vol. 1    (J. Wiley and Sons, 2007).
18  Sachdev, S. *Quantum Phase Transitions*.  (Cambridge University Press, Cambridge, U.K., 1999).
19  Varma, C. M. Mixed-valence compounds. *Reviews of Modern Physics* **48**, 219-238 (1976).
20  Grenzebach, C., Anders, F. B., Czycholl, G. & Pruschke, T. Transport properties of heavy-fermion systems. *Phys. Rev. B* **74**, 195119 (2006).
21  Shim, J. H., Haule, K. & Kotliar, G. Modeling the Localized-to-Itinerant Electronic Transition in the Heavy Fermion System $CeIrIn_5$. *Science* **318**, 1615-1617 (2007).
22  Martin, L. C., Bercx, M. & Assaad, F. F. Fermi surface topology of the two-dimensional Kondo lattice model: Dynamical cluster approximation approach. *Phys. Rev. B* **82**, 245105 (2010).

23. Jacob, D., Haule, K. & Kotliar, G. Dynamical mean-field theory for molecular electronics: Electronic structure and transport properties. *Phys. Rev. B* **82**, 195115 (2010).
24. Benlagra, A., Pruschke, T. & Vojta, M. Finite-temperature spectra and quasiparticle interference in Kondo lattices: From light electrons to coherent heavy quasiparticles. *Phys. Rev. B* **84**, 195141 (2011).
25. Yang, Y.-f. Fano effect in the point contact spectroscopy of heavy-electron materials. *Phys. Rev. B* **79**, 241107 (2009).
26. Maltseva, M., Dzero, M. & Coleman, P. Electron Cotunneling into a Kondo Lattice. *Phys. Rev. Lett.* **103**, 206402 (2009).
27. Figgins, J. & Morr, D. K. Differential Conductance and Quantum Interference in Kondo Systems. *Phys. Rev. Lett.* **104**, 187202 (2010).
28. Wölfle, P., Dubi, Y. & Balatsky, A. V. Tunneling into Clean Heavy Fermion Compounds: Origin of the Fano Line Shape. *Phys. Rev. Lett.* **105**, 246401 (2010).
29. Park, W. K., Sarrao, J. L., Thompson, J. D. & Greene, L. H. Andreev Reflection in Heavy-Fermion Superconductors and Order Parameter Symmetry in CeCoIn$_5$. *Phys. Rev. Lett.* **100**, 177001 (2008).
30. Aynajian, P. *et al.* Visualizing the formation of the Kondo lattice and the hidden order in URu$_2$Si$_2$. *Proc. Natl. Acad. Sci. U.S.A.* **107**, 10383 (2010).
31. Schmidt, A. R. *et al.* Imaging the Fano lattice to 'hidden order' transition in URu$_2$Si$_2$. *Nature* **465**, 570-576 (2010).
32. Ernst, S. *et al.* Emerging local Kondo screening and spatial coherence in the heavy-fermion metal YbRh$_2$Si$_2$. *Nature* **474**, 362-366 (2011).
33. Sidorov, V. A. *et al.* Superconductivity and Quantum Criticality in CeCoIn$_5$. *Phys. Rev. Lett.* **89**, 157004 (2002).
34. Paglione, J. *et al.* Field-Induced Quantum Critical Point in CeCoIn$_5$. *Phys. Rev. Lett.* **91**, 246405 (2003).
35. Paglione, J., Sayles, T. A., Ho, P. C., Jeffries, J. R. & Maple, M. B. Incoherent non-Fermi-liquid scattering in a Kondo lattice. *Nat. Phys.* **3**, 703-706 (2007).
36. Urbano, R. R. *et al.* Interacting Antiferromagnetic Droplets in Quantum Critical CeCoIn$_5$. *Phys. Rev. Lett.* **99**, 146402 (2007).
37. Hegger, H. *et al.* Pressure-Induced Superconductivity in Quasi-2D CeRhIn$_5$. *Phys. Rev. Lett.* **84**, 4986 (2000).
38. Petrovic, C. *et al.* Heavy-fermion superconductivity in CeCoIn$_5$ at 2.3 K. *J. Phys. Cond. Matter* **13**, L337-L342 (2001).
39. Hall, D. *et al.* Fermi surface of the heavy-fermion superconductor CeCoIn$_5$: The de Haas–van Alphen effect in the normal state. *Phys. Rev. B* **64**, 212508 (2001).
40. Shishido, H., Settai, R., Hashimoto, S., Inada, Y. & Ōnuki, Y. De Hass van Alphen effect of CeRhIn$_5$ and CeCoIn$_5$ under pressure. *J. Magn. Magn. Mater.* **272-276, Part 1**, 225-226 (2004).
41. Moshopoulou, E. G. *et al.* Comparison of the crystal structure of the heavy-fermion materials CeCoIn$_5$ , CeRhIn$_5$ and CeIrIn$_5$. *Appl. Phys. A* **74**, s895-s897 (2002).
42. Booth, C. H. *et al.* Local structure and site occupancy of Cd and Hg substitutions in CeTIn$_5$ (T=Co, Rh, and Ir). *Phys. Rev. B* **79**, 144519 (2009).
43. Ikeda, S. *et al.* Direct observation of a quasiparticle band in CeIrIn$_5$ : An angle-resolved photoemission spectroscopy study. *Phys. Rev. B* **73**, 224517 (2006).

44    Ehm, D. *et al.* High-resolution photoemission study on low- $T_K$ Ce systems: Kondo resonance, crystal field structures, and their temperature dependence. *Phys. Rev. B* **76**, 045117 (2007).

45    Koitzsch, A. *et al.* Hybridization effects in $CeCoIn_5$ observed by angle-resolved photoemission. *Phys. Rev. B* **77**, 155128 (2008).

46    Oppeneer, P. M. *et al.* Fermi surface changes due to localized–delocalized f-state transitions in Ce-115 and Pu-115 compounds. *J. Magn. Magn. Mater.* **310**, 1684-1690 (2007).

47    Sachdev, S. & Ye, J. Universal quantum-critical dynamics of two-dimensional antiferromagnets. *Phys. Rev. Lett.* **69**, 2411-2414 (1992).

48    Senthil, T. Critical Fermi surfaces and non-Fermi liquid metals. *Physical Review B* **78**, 035103 (2008).

49    Aronson, M. C. *et al.* Non-Fermi-Liquid Scaling of the Magnetic Response in $UCu_{5-x}Pd_x$ (x=1,1.5). *Phys. Rev. Lett.* **75**, 725-728 (1995).

50    Schröder, A., Aeppli, G., Bucher, E., Ramazashvili, R. & Coleman, P. Scaling of Magnetic Fluctuations near a Quantum Phase Transition. *Phys. Rev. Lett.* **80**, 5623-5626 (1998).

**Acknowledgements**

We gratefully acknowledge discussions with P. W. Anderson, E. Abrahams, P. Coleman, N. Curro, D. Pines, D. Morr, T. Senthil, S. Sachdev, M. Vojta, C. Varma, and C. V. Parker.

Work at Princeton was primarily supported by grant from DOE-BES. The instrumentation and infrastructure at the Princeton Nanoscale Microscopy Laboratory are also supported by grants from NSF-DMR, NSF-MRSEC program through Princeton Center for Complex Materials, and W.M. Keck foundation. P.A. acknowledges postdoctoral fellowship support through the Princeton Center for Complex Materials funded by the National Science Foundation MRSEC program.

Work at Los Alamos was performed under the auspices of the U.S. Department of Energy, Office of Basic Energy Sciences, Division of Materials Science and Engineering. Z. F. acknowledges support from NSF-DMR-0801253.

**Author Contributions** P.A., E.H.S.N., and A.G. performed the STM measurements. P.A. and E.H.S.N. analyzed the data. E.H.S.N. and P.A. performed the theoretical calculations. R.E.B., J.D.T., Z.F., and E.D.B. synthesized and characterized the materials. A.Y., P.A., E.H.S.N. wrote the manuscript. All authors commented on the manuscript.

**Figure Captions**

**Figure 1 | Tunneling into a Kondo lattice.** Schematic representations of a single-impurity Kondo effect **a** and a Kondo lattice **b** illustrating the screening (hybridization) of the local moments (red arrows) by the itinerant conduction electrons (green arrows). **c** Schematic phase diagram of heavy-fermion systems where the electronic ground state can be tuned from antiferromagnetism (AFM) with localized *f*-moments to a heavy Fermi liquid (HFL) with itinerant *f*-electrons. At low temperatures, superconductivity (SC) sets in near the quantum critical point (QCP) from a non-Fermi liquid (NFL). **d** Bare electronic bands (dashed lines) and hybridized heavy fermion bands (HF) (solid lines) displaying a direct ($2v$) and an indirect ($\Delta_h$) hybridization gaps. **e** Tunneling spectra computed from the hybridized band structure in **d** for a tunneling ratio $t_f/t_c$ = -0.025 showing sensitivity to the direct hybridization gap ($2v$). **f** Similar spectra computed with $t_f/t_c$ = -0.37 showing sensitivity to the indirect gap ($\Delta_h$). See Supplementary section I for details of the model.

**Figure 2 | STM topographies on CeCoIn$_5$. a** Constant current topographic image (+200 mV, 200 pA) showing an atomically ordered surface (termed surface A) with a lattice constant of ~ 4.6 Å. **b** Topographic image (-200 mV, 200 pA) showing two consecutive layers: a distinct atomically ordered surface (termed surface B, lattice constant ~ 4.6 Å) and a reconstructed surface (termed surface C). **c** Constant current topographic image (-150 mV, 365 pA) displaying all three surfaces (the derivative of the topography is shown to enhance contrast). **d** A line cut through the different surfaces (solid line in **c**) showing the relative step heights compared to the bulk crystal structure. Insets in **a** and **b** show blow-ups (45 x 45 Å$^2$) of the three different surfaces.

**Figure 3 | Composite nature of heavy fermion excitations. a** Averaged tunneling spectra (-150 mV, 200 pA) measured on surface A of CeCo(In$_{0.9985}$Hg$_{0.0015}$)$_5$ for different temperatures (solid lines) and on the corresponding surface A of CeRhIn$_5$ at 20 K (dashed line). **b** Averaged tunneling spectra (-150 mV, 200 pA) measured on surface B of CeCo(In$_{0.9985}$Hg$_{0.0015}$)$_5$ for different temperatures (solid lines) and on corresponding surface B of CeRhIn$_5$ at 20 K (dashed line). **c, d** Tunneling spectra computed for $t_f/t_c$ = -0.01 (**c**) and $t_f/t_c$ = -0.20 (**d**) for selected values of $\gamma_f$. See Supplementary section I for details of the model.

**Figure 4 | Spectroscopic mapping of quasiparticle interference.** Real space **a** and corresponding DFT **b** of conductance maps (-200 mV, 1.6 nA) at selected energies measured on surface A of CeCo(In$_{0.9985}$Hg$_{0.0015}$)$_5$ at 20 K. Similar DFTs for CeCo(In$_{0.9985}$Hg$_{0.0015}$)$_5$ at 70 K (-150 mV, 1.5 nA) **c** and on the corresponding surface A for CeRhIn$_5$ at 20 K (-200mV, 3.0 nA) **d** at selected energies. The arrow indicates the position of the Bragg peaks at ($2\pi/a$, 0) and (0, $2\pi/a$). All DFTs were four-fold symmetrized (due to

the four-fold crystal symmetry) to enhance resolution (see Supplementary section IV). The intensity is represented on a linear scale.

**Figure 5 | Visualizing quasiparticle mass enhancement.** Energy-momentum structure of the QPI bands extracted from line cuts (solid white lines in Fig. 4b) along the atomic direction ($2\pi/a$, 0) **a** and along the zone diagonal ($\pi/a$, $\pi/a$) **b** in CeCo(In$_{0.9985}$Hg$_{0.0015}$)$_5$ at 20 K. The solid red line represents a fourth-order polynomial fit to the data. Inset in **a** shows the effective mass $m^*/m_0$ as a function of momenta obtained from the curvature (¼ $\hbar^2 [d^2E/dq^2]^{-1}$) of the outer band (solid red line in **a**). **c** Average spectrum on surface A of CeCo(In$_{0.9985}$Hg$_{0.0015}$)$_5$ at 20 K reflecting the suppression of scattering in the QPI bands. Similar measurements performed in CeCo(In$_{0.9985}$Hg$_{0.0015}$)$_5$ at 70 K **d**, **e**, **f** and in CeRhIn$_5$ at 20 K **g**, **h**, **i** .PSD, power spectral density. The intensity is represented on a linear scale.

**Figure 6 | Signatures of quantum criticality. a** Full width at half maximum (FWHM) of the heavy quasiparticle peak (red squares) as a function of temperature extracted from a Gaussian fit to the sharp lineshape of the spectra of Fig. 3b after a smooth background subtraction (see supplementary section VI). Blue squares represent the thermally deconvoluted FWHM corresponding to the intrinsic width in the absence of thermal broadening. The green line represents $3/2k_BT$. The error bars represent one standard deviation. **b** Energy-temperature scaling of the different spectra of Fig. 3b, after the removal of a temperature independent background (see supplementary section VI and VII), within a narrow energy window near the Fermi energy. Inset shows the 'goodness of the collapse' as a function of the critical exponents $\alpha$ and $\beta$.

**Figure 1**

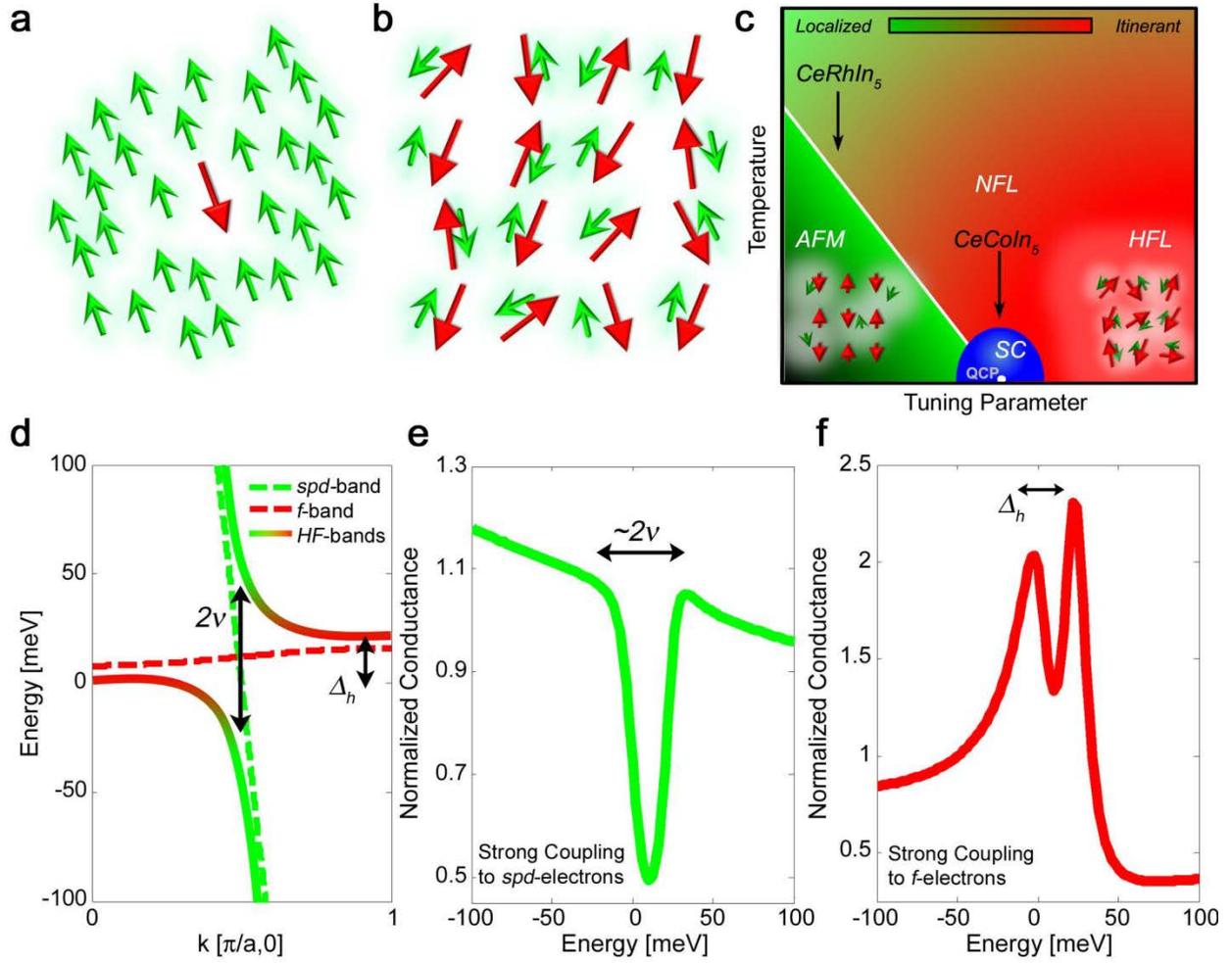

**Figure 2**

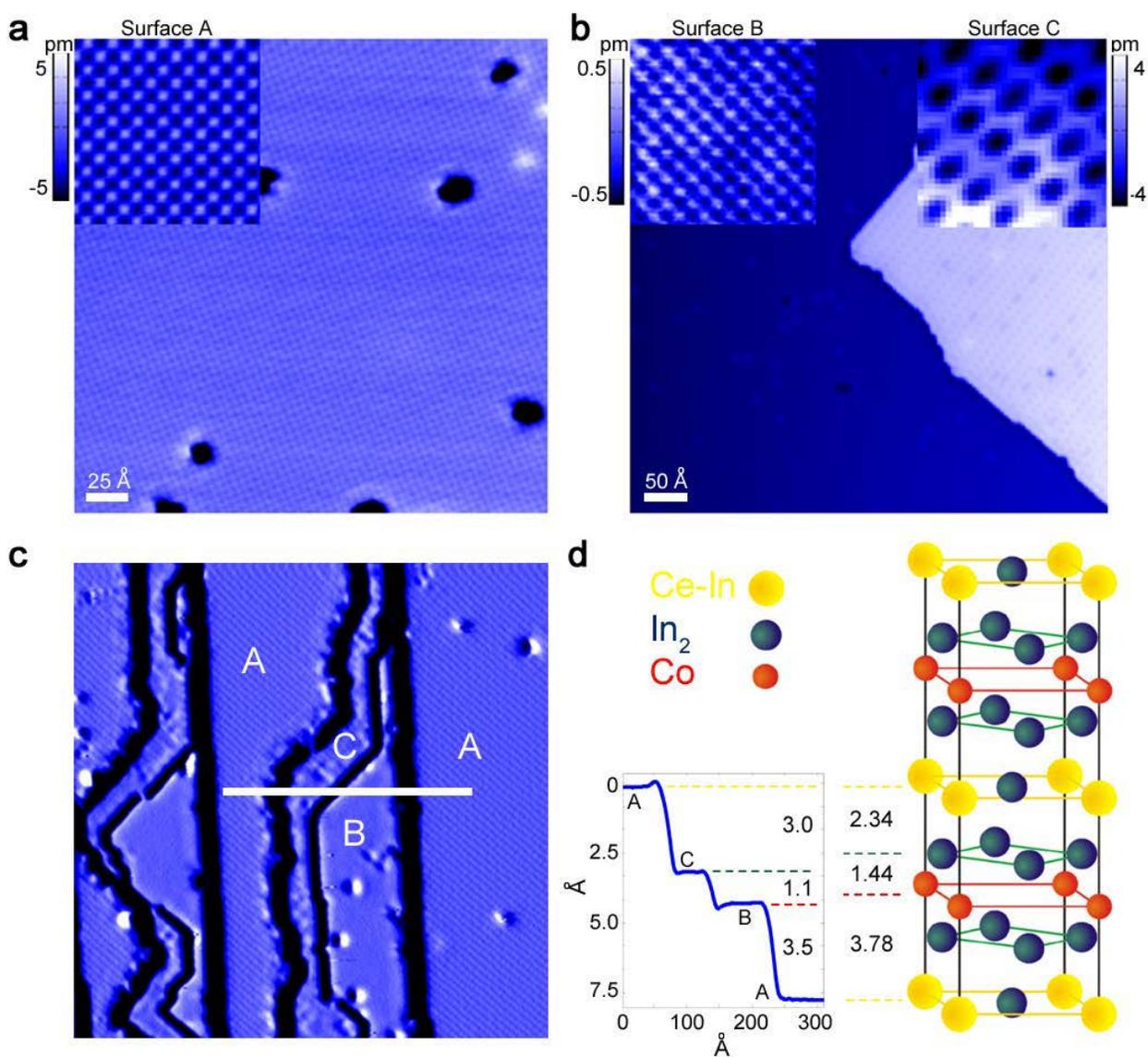

**Figure 3**

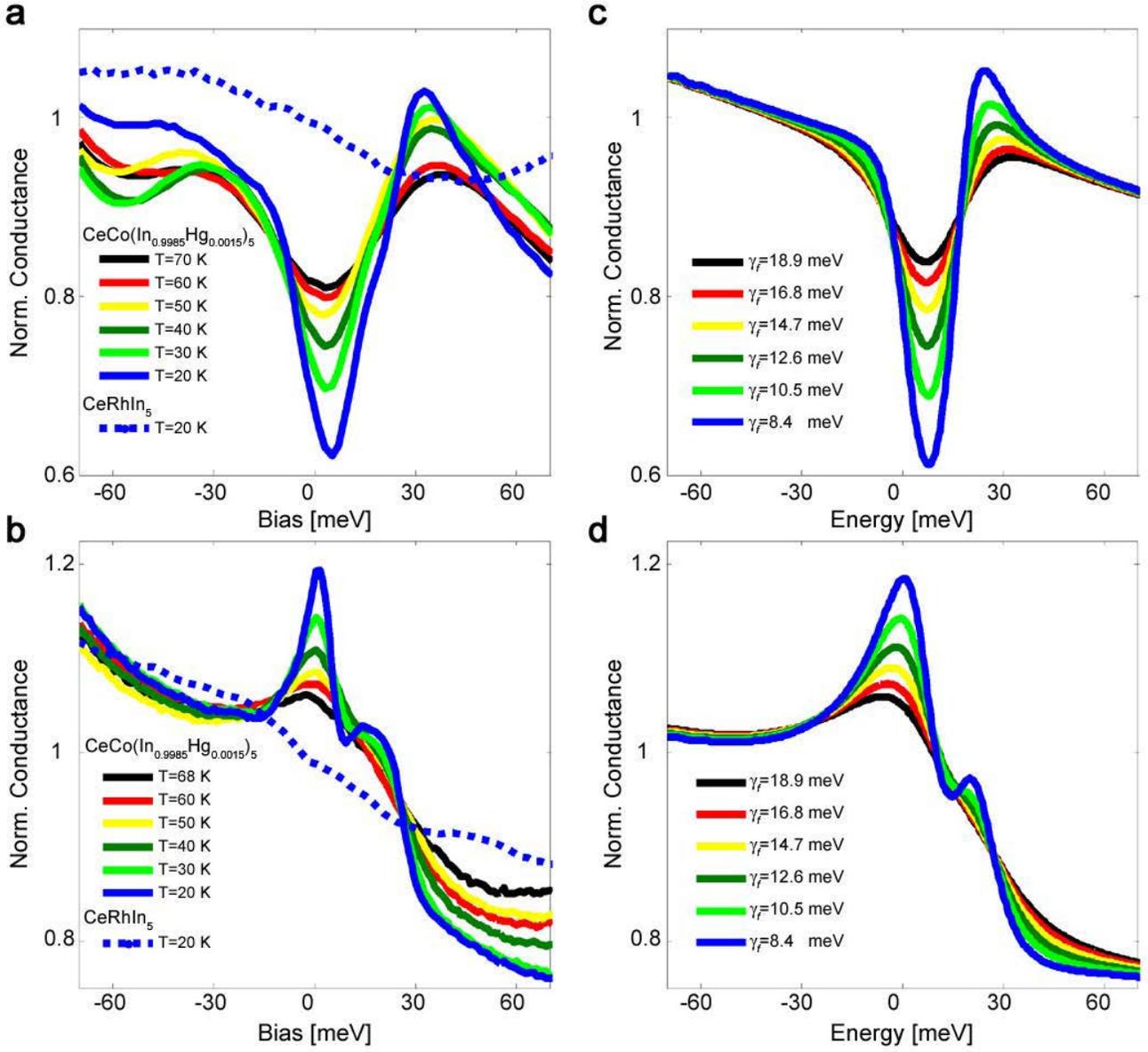

**Figure 4**

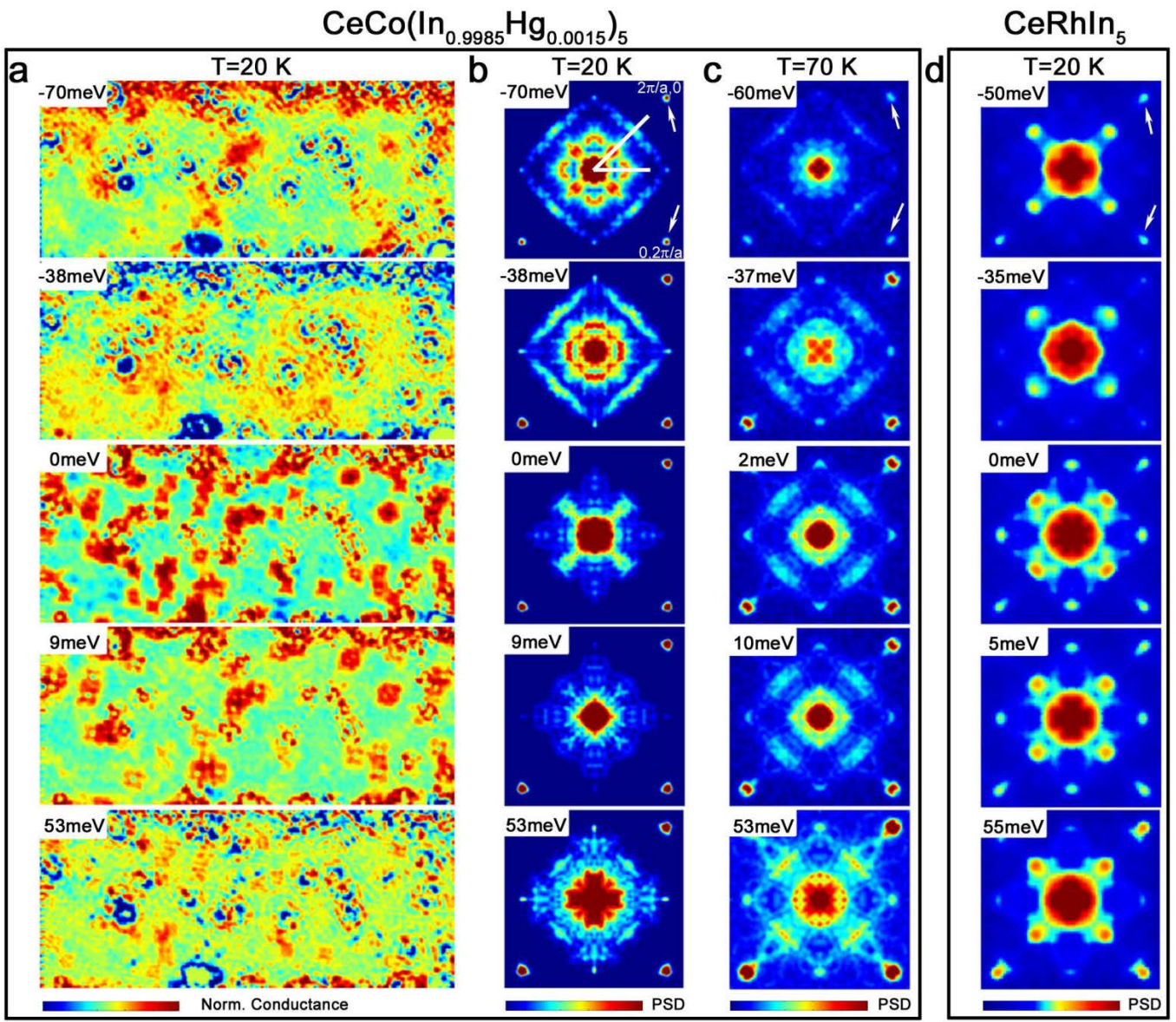

**Figure 5**

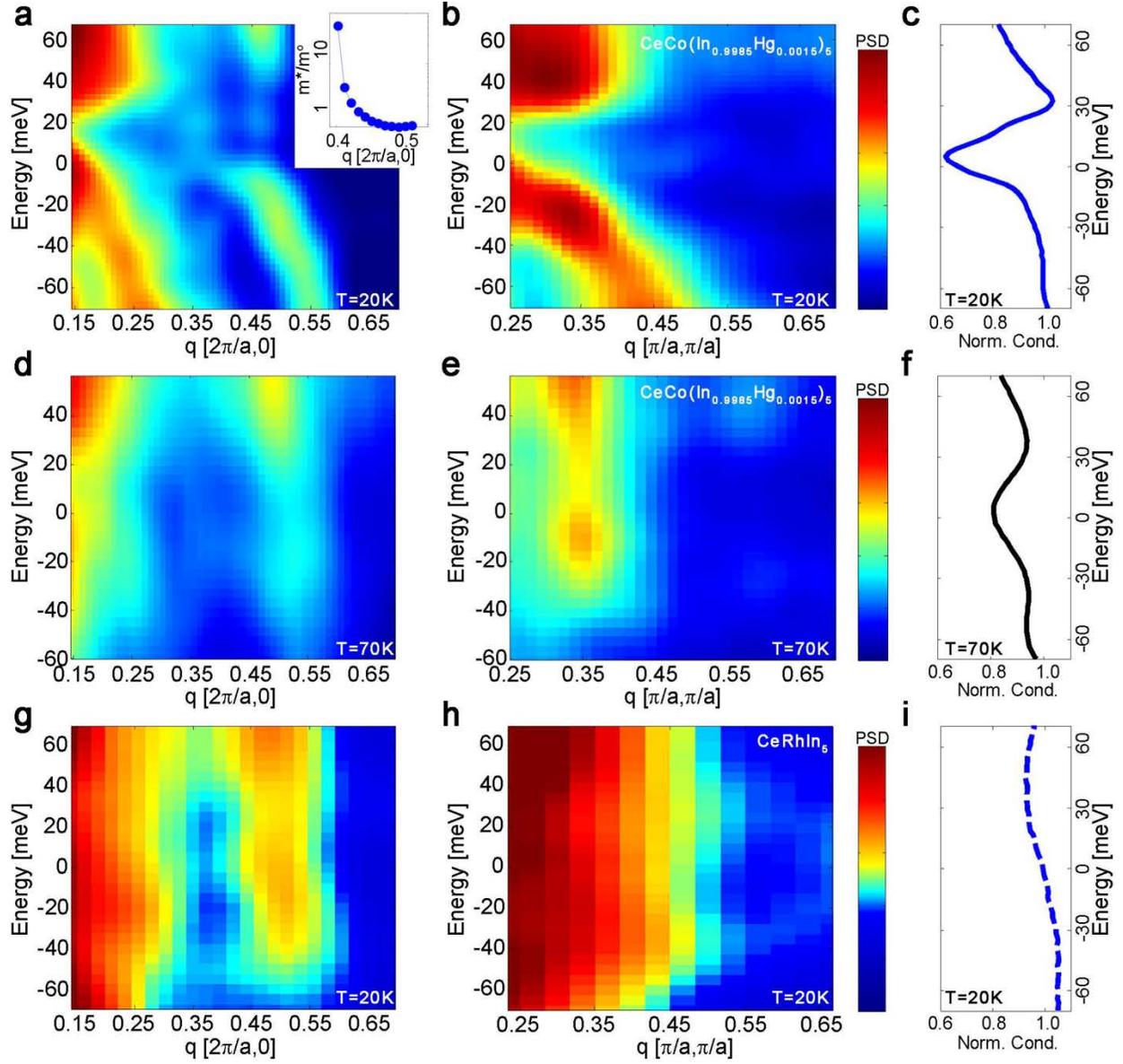

**Figure 6**

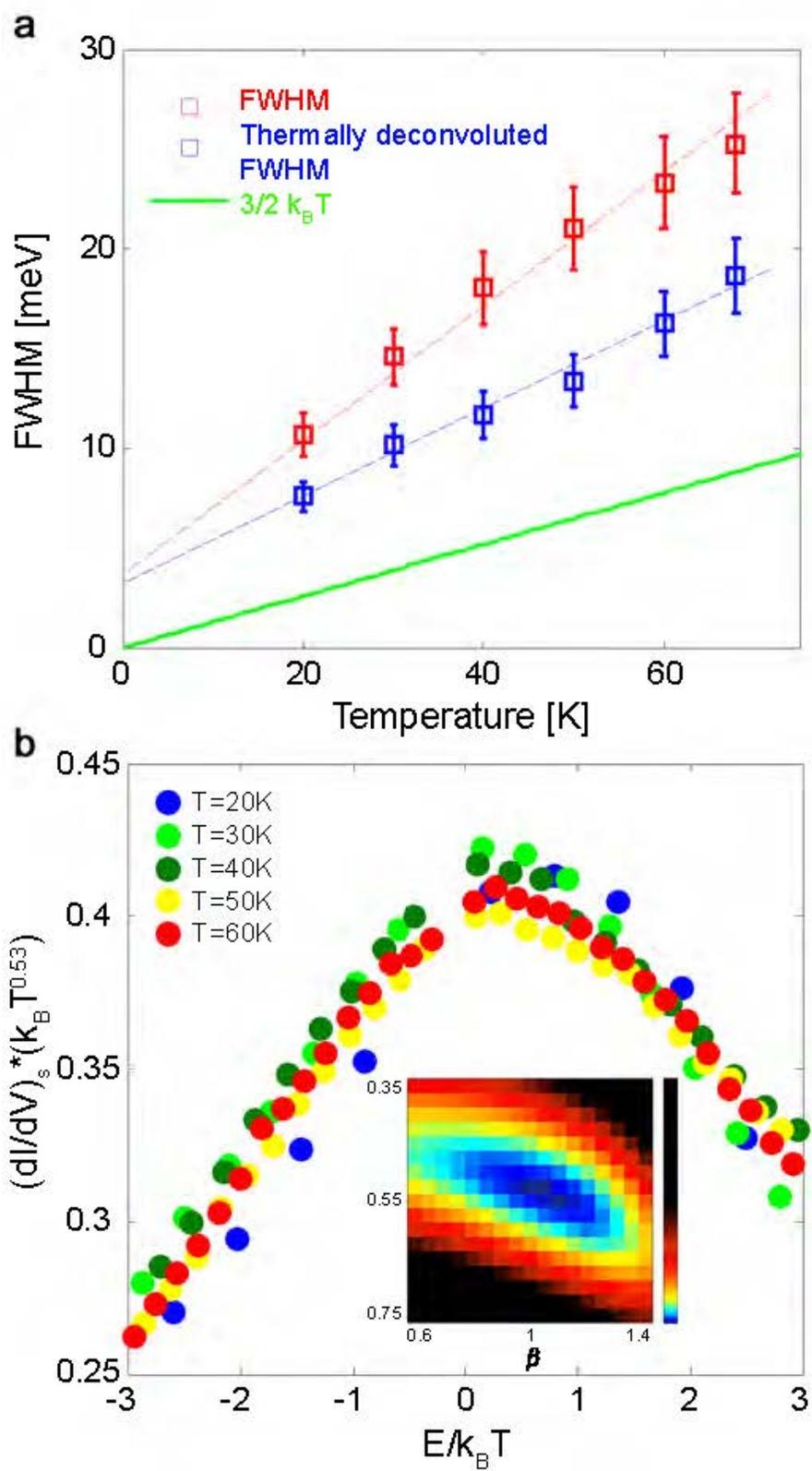

**Supplementary Information for "Visualizing heavy fermions emerging in a quantum critical Kondo lattice"**

Pegor Aynajian[1]*, Eduardo H. da Silva Neto[1]*, András Gyenis[1], Ryan E. Baumbach[2], J. D. Thompson[2], Zachary Fisk[3], Eric D. Bauer[2], and Ali Yazdani[1]§

[1]Joseph Henry Laboratories and Department of Physics, Princeton University, Princeton, NJ 08544 USA
[2]Los Alamos National Laboratory, Los Alamos, New Mexico 87545, USA
[3]University of California, Irvine, California 92697, USA

§To whom correspondence should be addressed. Email: yazdani@princeton.edu
*These authors contributed equally to this work.

### I. Modeling the tunneling density of states

To understand experimentally observed spectroscopic lineshapes (*dI/dV*) and the quasiparticle interference (QPI) patterns and show that they are consistent with the emergence of heavy fermions, we model the *dI/dV* following a recent theoretical description of a coherent Kondo lattice state described in ref. (S1). The model describes the formation of heavy fermions using the Kondo-Heisenberg Hamiltonian within the framework of a large-N approach and computes the differential conductance *dI/dV* measured in STM experiments. The significance of the model comes from the introduction of two different tunneling amplitudes ($t_f$ and $t_c$) to the two different electronic states (heavy *f*-band and light conduction band). The ratio between these tunneling paths and their interference determines the experimental lineshapes of the spectra.

The starting point is a light hole-like conduction (*c*) band $E_k^c = 2t(\cos k_x + \cos k_y) - \mu$ and a heavy flat (*f*) band $E_k^f = -2\chi_0(\cos k_x + \cos k_y) - 4\chi_1 \cos k_x \cos k_y + \varepsilon_o^f$ near the Fermi energy, where $t$ and $\mu$ represent the nearest neighbor hopping of the conduction electrons and the chemical potential, respectively, and $\chi_0$, $\chi_1$, and $\varepsilon_o^f$ represent the nearest and next-nearest site spin correlations, and the position of the heavy band with respect to the Fermi energy, respectively. The differential conductance *dI/dV* is given by

$$dI(k,\omega)/dV = -\frac{2e}{\hbar}\hat{T}\sum_{i,j=1}^{2}\left[\hat{t}\,\text{Im}\hat{G}(k,\omega)\hat{t}\right]_{ij},$$

Where the matrix $\hat{t} = \begin{pmatrix} t_c & 0 \\ 0 & t_f \end{pmatrix}$ controls the ratio of tunneling to the c- and f- bands, $\hat{T} = 1$ is the STM tip density of states, and $\hat{G} = \begin{pmatrix} G_{cc} & G_{cf} \\ G_{fc} & G_{ff} \end{pmatrix}$ defines the full Green's function describing the hybridization between the c- and f-electron bands, with

$$G_{ff}(k,\omega) = \left[(G_{ff}^0(k,\omega))^{-1} - v^2 G_{cc}^0(k,\omega)\right]^{-1}$$

$$G_{cc}(k,\omega) = \left[(G_{cc}^0(k,\omega))^{-1} - v^2 G_{ff}^0(k,\omega)\right]^{-1}$$

$$G_{cf}(k,\omega) = G_{cc}^0(k,\omega) v G_{ff}(k,\omega),$$

where $G_{cc}^0(k,\omega) = (\omega + i\gamma_c - E_k^c)^{-1}$ and $G_{ff}^0(k,\omega) = (\omega + i\gamma_f - E_k^f)^{-1}$ are the bare Green's function and $\gamma_c^{-1}$ and $\gamma_f^{-1}$ are the lifetimes of the c- and f-electron states, respectively. The poles of the above Green's function yields two heavy fermion bands $E_k^\pm$:

$$E_k^\pm = \frac{E_k^c + E_k^f}{2} \pm \sqrt{\left(\frac{E_k^c - E_k^f}{2}\right)^2 + v^2}$$

with $v$ being the hybridization amplitude between the c (light) and f (heavy) bands.

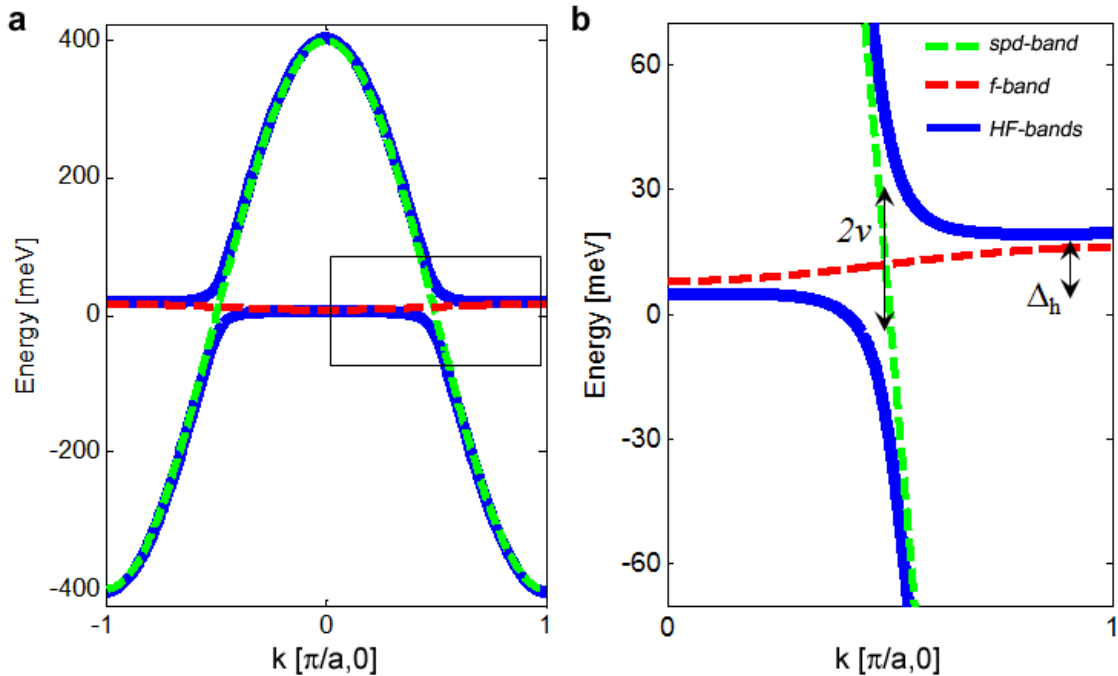

**Fig. S1**: **a** Dispersion of the bare heavy (dashed red) and light (dashed green) electronic bands and the hybridized heavy fermion bands (solid blue) computed for $t = 200\,\text{meV}$, $\mu = 2.00t$, $\chi_0 = 0.01t$, $\chi_1 = 0.01\chi_0$, $\varepsilon_o^f = 0.08t$, and $v = 0.18t$. **b** Blow-up of the square area in **a**.

Figure S1 shows a plot of the bare heavy and light electronic bands (dotted lines) and the hybridized heavy fermion bands (solid lines) and Fig. S2 displays the computed differential conductance using the same band structure of Fig. S1 for several selected values of $t_f/t_c$. For small $t_f/t_c$, the predominant tunneling channel is into the light conduction band and the differential conductance exhibits a gap. For larger $t_f/t_c$, tunneling becomes more sensitive to the f-orbitals and the differential conductance exhibits two sharp peaks reflecting the electronic density of states of the split heavy fermion bands and the opening of the small indirect hybridization gap $\Delta_h$. To generate the spectra in Figs. 3c,d (of the main text) the band structure of Fig. S1 was used for $\gamma_c = 0.03t$ and for different values of $\gamma_f$.

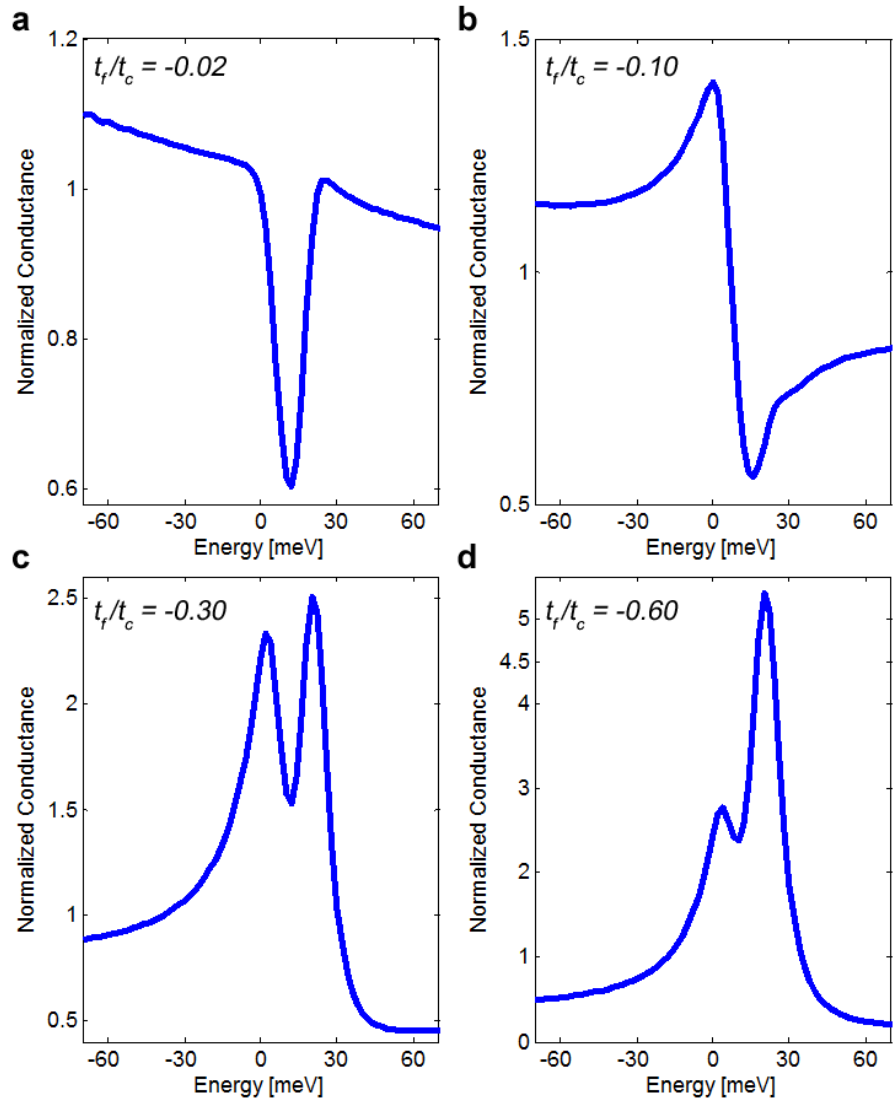

**Fig. S2**: Differential conductance computed using the same band structure parameters as in Fig. S1 for selected values of $t_f/t_c$. The inverse quasiparticle lifetimes used are $\gamma_c = 0.03t$ and $\gamma_f = 0.03t$.

From our model, even in the absence of any lifetime broadening $\gamma_f$, the peaks in the simulated spectra for large values of $t_f/t_c$ still exhibit a finite width. This is caused by the finite dispersion of the *f*-band (unrenormalized *f* bandwidth of 8 meV, see Fig. S1b) and the finite value of $t_f/t_c$. The simulation presented in Fig. S3 addresses this point for an infinitesimally small $\gamma_f$. Therefore at zero temperature and for infinitesimally small scattering, the STM measurements such as that in Figure 3a of the main manuscript would still have a finite width. The zero temperature intercept extracted from such data in Figure 6a is likely the results of such behavior.

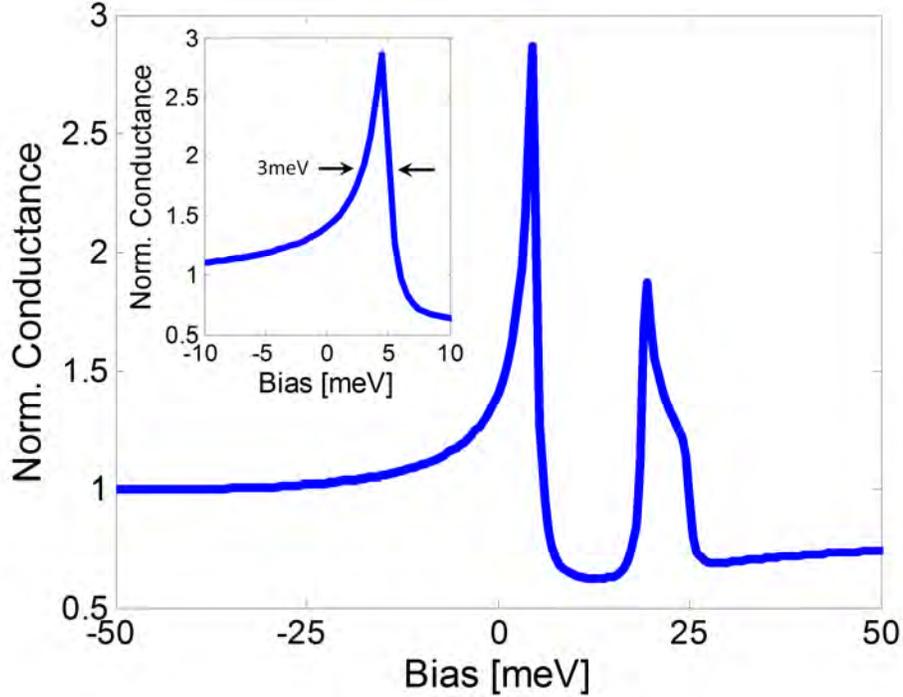

**Fig. S3**: Differential conductance computed using the same band structure parameters as in Fig. 3d of the main text for $\gamma_f = 0.0025t$, corresponding to a 0.5 meV inverse quasiparticle lifetime broadening. Note that the two peaks are much broader than $\gamma_f$ and their widths are due to the finite *f*-band dispersion and the finite value of $t_f/t_c$ =-0.2 used.

To speculate the quasiparticle interference data, the Fourier transform of the *dI/dV* into *q*-space is computed:

$$S(q,\omega) = dI(q,\omega)/dV = 2\pi e/\hbar \hat{T} \sum_{i,j=1}^{2} \left[\hat{t}\hat{N}(q,\omega)\hat{t}\right]_{ij}$$

Where $\hat{N}(q,\omega) = -\frac{1}{\pi} \text{Im} \int \frac{d^2k}{(2\pi)^2} \hat{G}(k,\omega)\hat{U}\hat{G}(k+q,\omega)$ represents the single impurity scattering between points on the contours of constant energy of the hybridized band structure and $\hat{U} = \begin{pmatrix} U_c & 0 \\ 0 & U_f \end{pmatrix}$, with $U_c$ and $U_f$ being the scattering potential in the $c$- and $f$-electron channels.

Figure S4 shows $S(q,\omega)$ computed for a scattering potential of $U_c = 1$ and $U_f = 0$ for two values of $\gamma_f$ showing the momentum-energy signatures of the hybridization. This reproduces the salient features of the heavy quasiparticles in the experimental data on CeCo(In$_{0.9985}$Hg$_{0.0015}$)$_5$ (Fig. 5a,b main text). Finally, by varying the coherence of the $f$-band we can reproduce the disappearance of the signatures of heavy quasiparticles in the data at high temperature (Fig. 5d,e main text).

While the simplified model presented here (which takes into account only a single conduction band) reproduces the main experimental spectroscopic features (dI/dV and QPI) remarkably well, a detailed analysis including the full LDA Green's function is essential to capture the finer details of both the spectroscopic lineshapes and the multiple QPI bands.

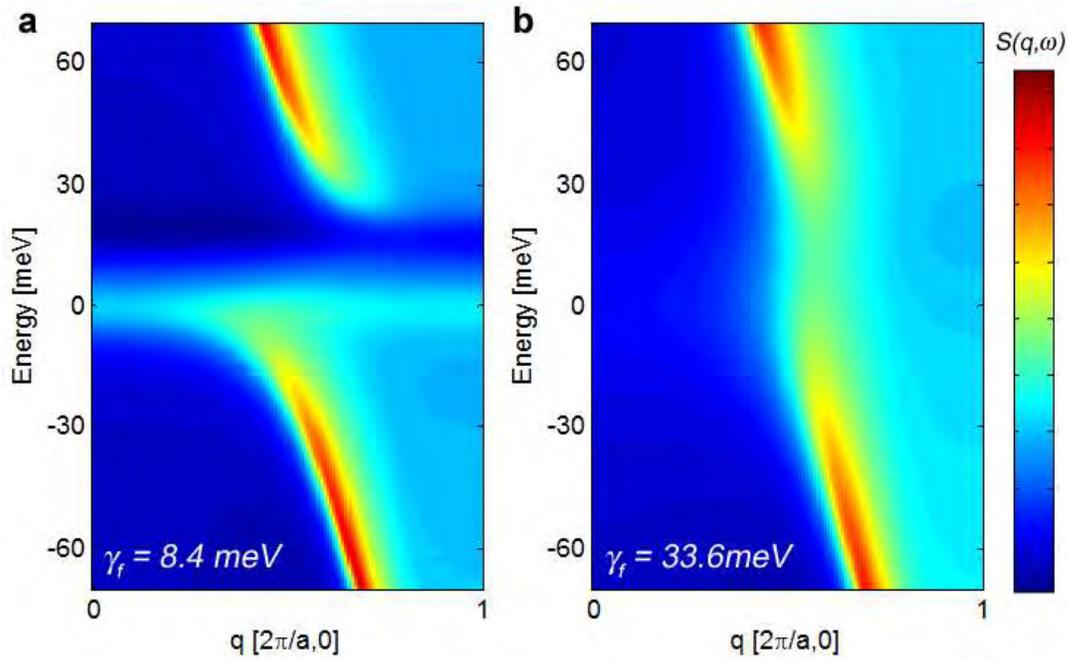

**Fig. S4**: The QPI band structure $S(q,\omega)$ for scattering potential of $U_c = 1$ and $U_f = 0$ computed from the same band structure of Fig. S1 for two selected values of $\gamma_f$.

## II. The cleavage planes in Ce-115 compounds

Cleaving of the three different Ce-115 compounds $CeCoIn_5$, $CeCo(In_{0.9985}Hg_{0.0015})_5$, and $CeRhIn_5$, revealed identical sets of cleavage planes. In all cases, multiple surfaces were obtained after cleaving with similar sub-unit cell step heights and surface topographies. Figure S5 shows the multiple surfaces and their relative step heights in $CeRhIn_5$ and compares the exposed layers to the bulk crystal structure. The exposed surfaces A', B', and C' are identified as Ce-In, Rh, and $In_2$, respectively in correspondence with $CeCoIn_5$ (Fig. 2 of main text).

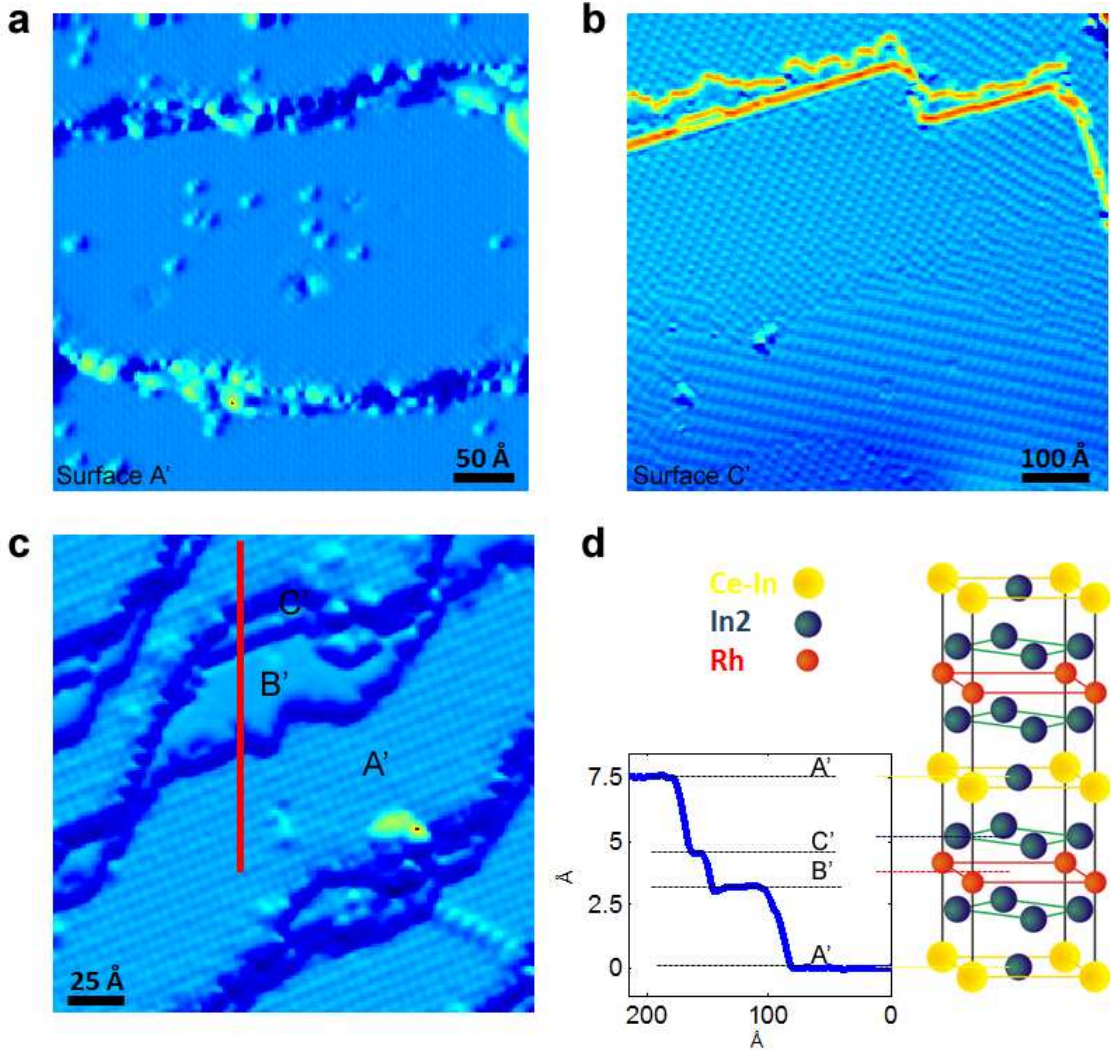

**Fig. S5**: STM topography of the different exposed surfaces in $CeRhIn_5$. **a** The atomic surface A' corresponding to surface A of $CeCoIn_5$. **b** The reconstructed surface C' corresponding to surface C of $CeCoIn_5$. Note the unidirectional electronic structure on the reconstructed surface which follows structural defects and step edges. **c** Relative sub-unit cell step heights within the different layers. **d** A line cut through the different surfaces (solid line in **c**) showing the relative step heights compared to the bulk crystal structure.

Figure S6 displays a topographic image of the A surface in CeCo(In$_{0.9985}$Hg$_{0.0015}$)$_5$ showing the dopant Hg atoms which replace the surface In atoms. Figure S6a shows the topography of the area on which the conductance maps of Fig. 4a (of the main text) were measured.

In all samples the surfaces B (or B') and C (or C') were exposed with equal probabilities whereas, within the limited statistics, surface A (or A') was obtained with a slightly lower probability. In some cases, insulating or unstable surfaces were obtained which immediately destroyed the STM tip and terminated the experiment.

In principle, breaking of the two different chemical bonds between the different layers would result in four exposed surfaces, namely, Ce-In, In$_2$ on Co, Co, and In$_2$ on Ce-In. Within the limited statistics, the absence of the second In$_2$ surface (on Ce-In) might be associated with the observation of unstable insulating surfaces.

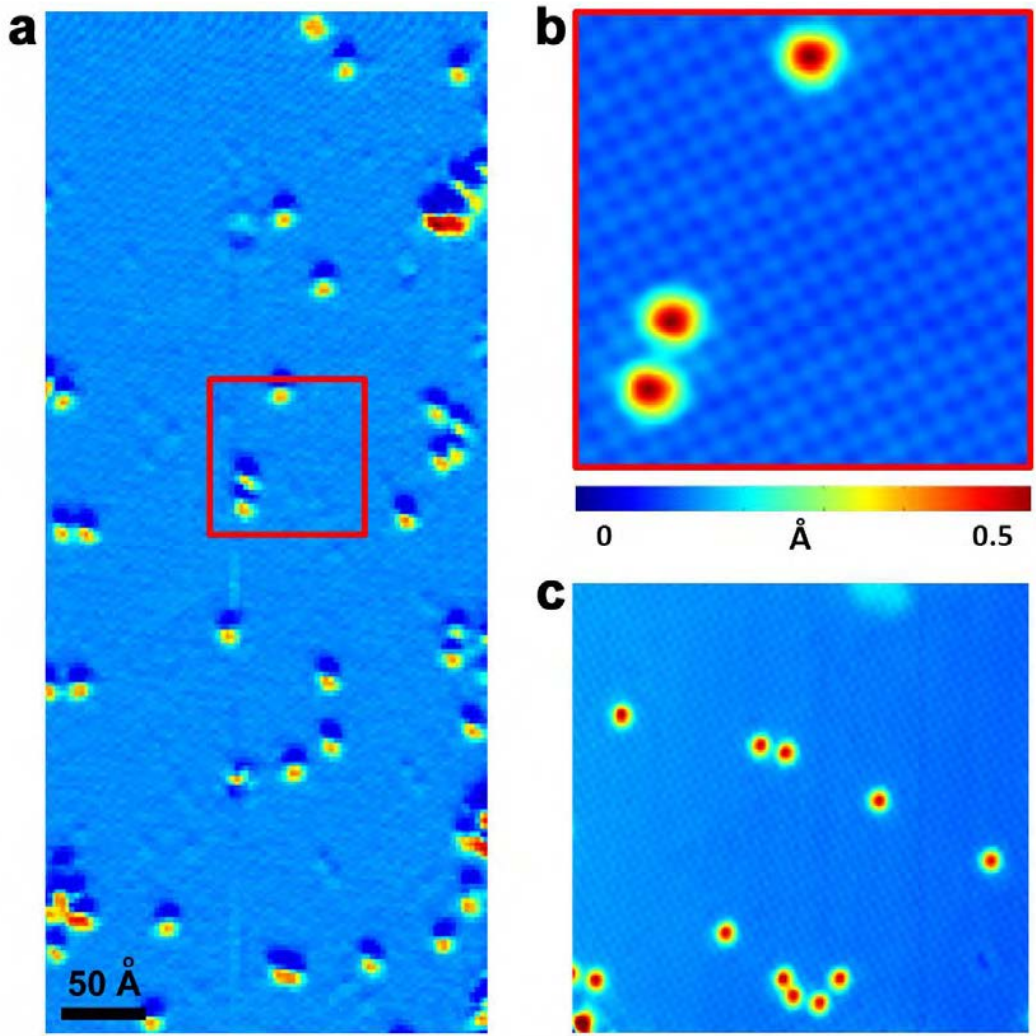

**Fig. S6**: Topographic image of the A surface in CeCo(In$_{0.9985}$Hg$_{0.0015}$)$_5$ showing the dopant Hg atoms which replace the surface In atoms.

### III. Spectroscopy on CeCoIn$_5$

Spectroscopic measurements were performed on both CeCoIn$_5$ and CeCo(In$_{0.9985}$Hg$_{0.0015}$)$_5$, the latter to enhance the QPI signal. Spectra on both samples revealed similar results, further verifying the negligible effect of 0.15% Hg doping to the transport properties. Fig. S7 displays temperature dependent spectra measured on surface A of an undoped CeCoIn$_5$ which shows the development of the hybridization gap very similar to the data on CeCo(In$_{0.9985}$Hg$_{0.0015}$)$_5$ presented in Fig. 3a of the main text.

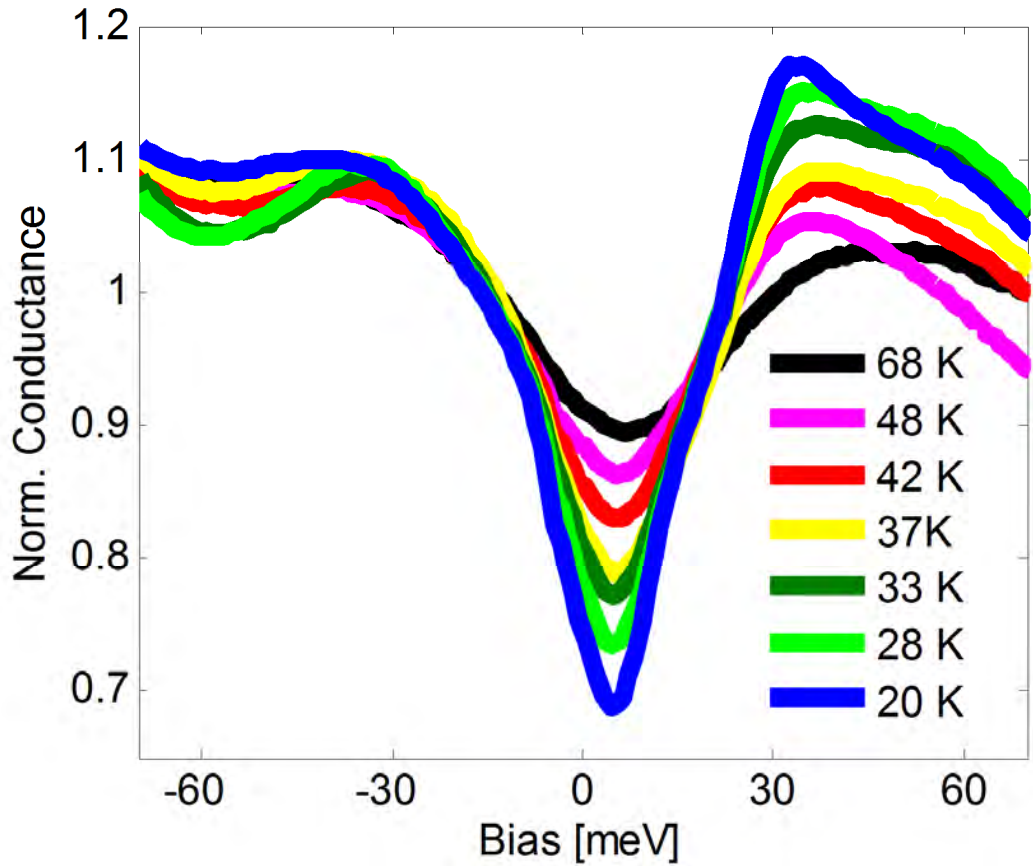

**Fig. S7**: STM spectroscopy measured on surface A of CeCoIn$_5$ as a function of temperature.

## IV. Raw and symmetrized QPI

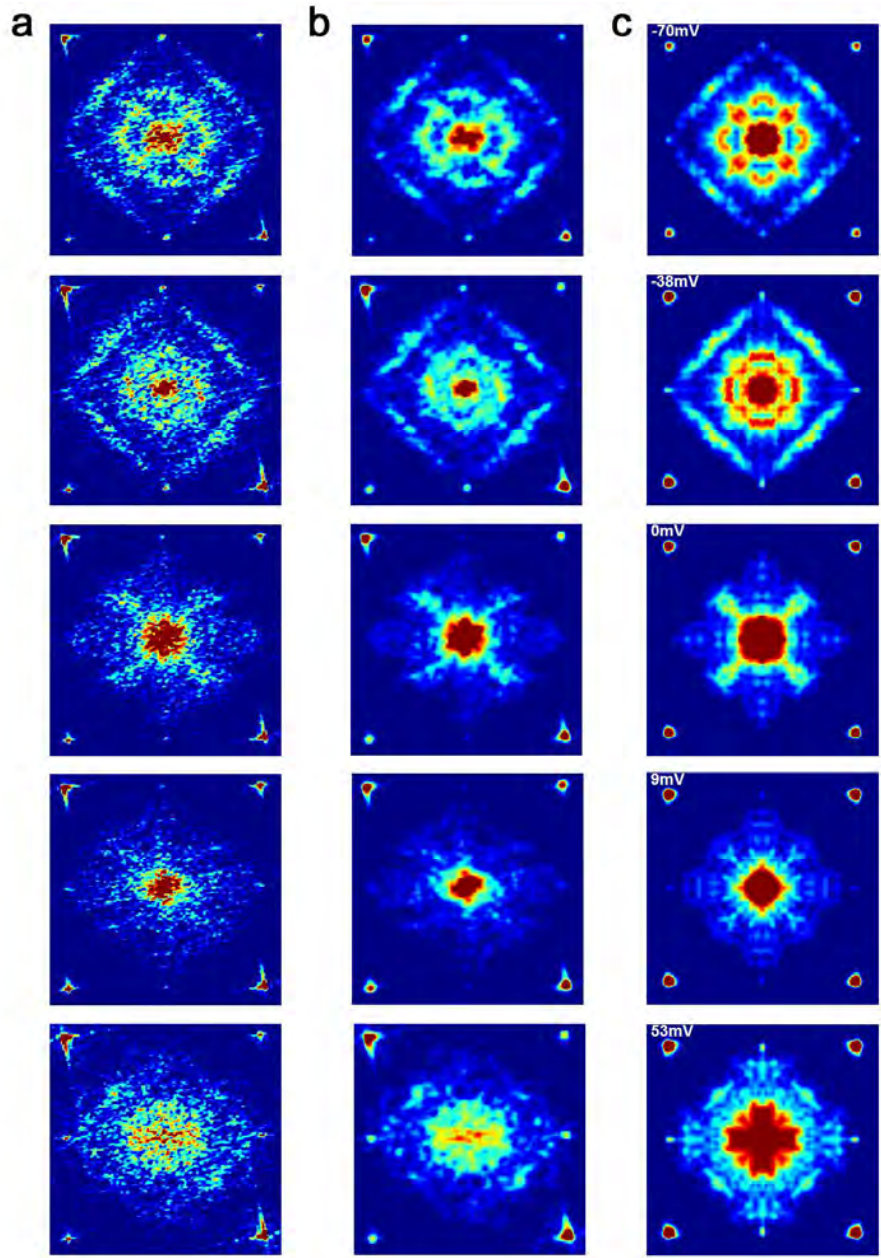

**Fig. S8**: **a** Raw, **b** Gaussian smoothened, and **c** Gaussian smoothened and four-fold symmetrized QPI data for CeCo(In$_{0.9985}$Hg$_{0.0015}$)$_5$ measured at 20K. Both raw and analyzed data show clearly the different scattering vectors.

## V. Comparing the QPI bands with the LDA band structure in Ce-115 compounds

A theoretical calculation of QPI from the full bulk band structure is required to understand the full details of our QPI measurements. However, in the absence of such calculations, one can still make progress understanding the QPI patterns by considering the two-dimensional QPI pattern observed in our experiments at the surface of this material as a 2D projection of the possible scattering within the different $k_z$ planes of the 3D Brillouin zone. We can make progress by combining this idea, with the fact that the most intense QPI signals will be dominated by scattering between points on the surfaces of constant energy in k-space with the highest density of states. Furthermore, one would expect that the most prominent features of the QPI signal measured at a surface would be related to wavevectors connecting nearly parallel contour (of constant energy) sheets. Following this approach, the square-like shape of the quasiparticle scattering wavevectors (*q* in Fig. S9a) can be understood as a result of the inter-band scattering of quasiparticles between the two conduction bands centered at the Brillouin zone corners (Points A in Fig. S9b,c) as calculated by LDA for both $CeRhIn_5$ and $CeCoIn_5$ (Fig. S9b,c) (S2, S3).

Figure S9c displays a cut of the LDA Fermi surface in the Z-plane of the Brillouin zone for $CeCoIn_5$ (S3) together with the magnitude of the QPI wavevector *q* (black arrows), which matches the inter-band separation of the two square-like bands (black arrows in Fig. S9c). While the Fermi surface exhibits multiple sheets, scattering is expected to be strongly enhanced between these two square-like bands owing to their two dimensional nature (Fig. S9b). The other features of the QPI with smaller wavevectors require more detailed analysis since they can originate from several combinations which connect different Fermi surface sheets.

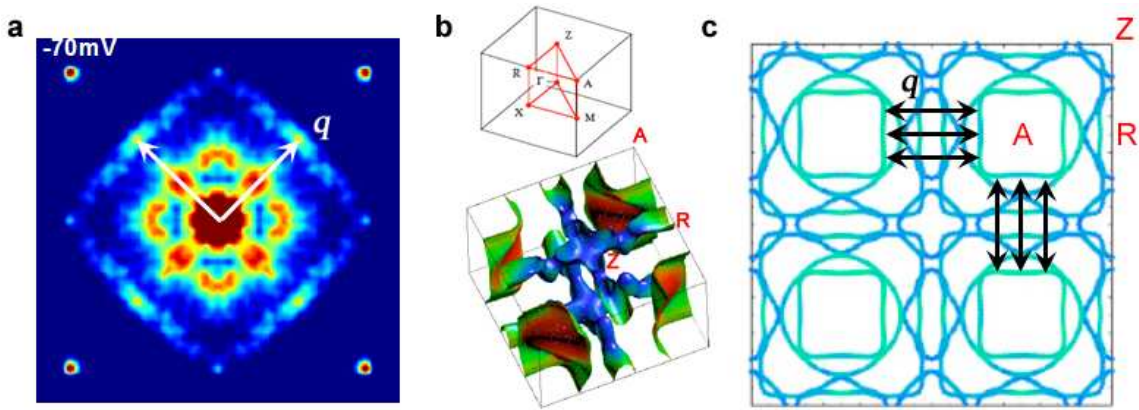

**Fig. S9**: **a** Quasiparticle interference pattern in $CeCo(In_{0.9985}Hg_{0.0015})_5$ displaying two concentric square like patterns. Arrows indicate the magnitude *q* of the outer quasiparticle band. **b** LDA calculation of the (135 sheet) Fermi surface in $CeCoIn_5$ displaying four cylindrical sheets at the corners of the Brillouin zone. **c** Cut of the Fermi surface in the Z-plane illustrating the sheets which are connected by the scattering wave vector *q* (black arrows).

**Transition from small to large Fermi surface**

Figure S10 shows the QPI band structure in CeCo(In$_{0.9985}$Hg$_{0.0015}$)$_5$ at 70 K and 20 K. At 70K the QPI band crosses the Fermi energy at a wave vector $k_F^{70}$ ~ 0.52 rlu (rlu: reciprocal lattice units). Heavy fermion hybridization at low temperatures alters the band structure splitting the band and forming two heavy fermion bands. At 20K the lower heavy fermion QPI band, due to this hybridization, cuts the Fermi surface at a smaller wave vector $k_F^{20}$ ~ 0.40 rlu. Owing to the fact that the QPI wavevectors are scattering vectors, therefore, can represent either direct or *Umklapp* scattering. In the latter case, decreasing QPI wavevector corresponds to an enlarging Fermi surface. Such a scenario is also consistent with the scattering analysis of the previous section where the wavevector q is an *Umklapp* scattering connecting the Fermi sheets across different Brillouin zones (Fig. S9c).

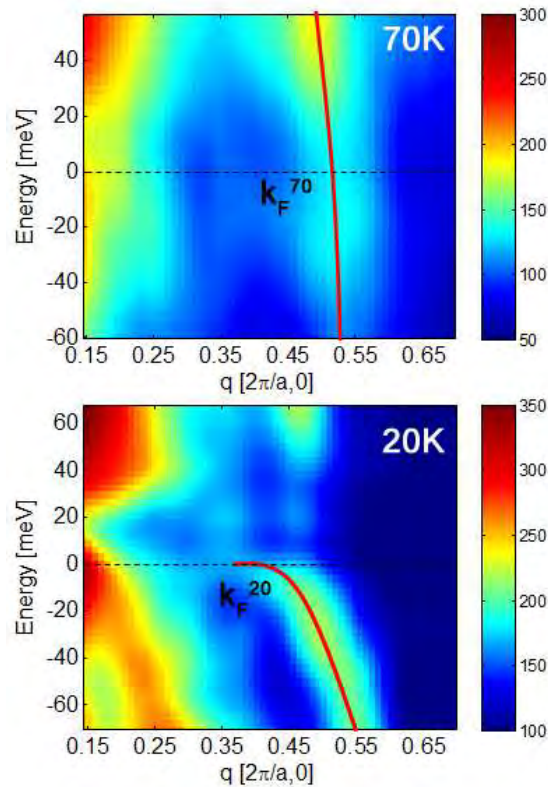

**Fig. S10:** Transition from a small to large Fermi surface. The crossing between the solid red (QPI band) and the dashed black (Fermi energy) lines define the QPI Fermi wavevector.

## VI. Heavy quasiparticle lifetimes and energy-temperature scaling

Information on the lifetime of heavy quasiparticles can be obtained by analyzing the widths of the sharp spectroscopic features in the lineshapes of Fig. 3b. In order to isolate the heavy quasiparticle peak from the rather smooth background, we fit a second order polynomial to the data outside the bias range of ± 50 meV (Fig. S11). This range of bias lies outside the hybridization energy scale where the density of states reflects the bare light electronic bands. Fig. S12 displays the background (polynomial) subtracted spectra $(dI/dV)_S$ for the different temperatures. The clear double peak structure near the Fermi energy, reflecting the heavy electronic states, is apparent. In addition to this double peak structure, a much broader and a weakly temperature dependent shoulder, near -20 meV, is also visible in the data, which could be related to crystal field excitations. To extract the linewidth of the sharp quasiparticle peak near the Fermi energy, the spectra are fitted to three Gaussian peaks centered near ~ -20, 1, and 18 meV. The extracted linewidth of the central peak near the Fermi energy, representing the inverse heavy quasiparticle lifetime, is displayed in Fig. 6a of the main text. The scaling in Fig. 6b are obtained from the data after subtracting the smooth background, following similar procedure such as those in measurements of dynamical susceptibility (S4,S5).

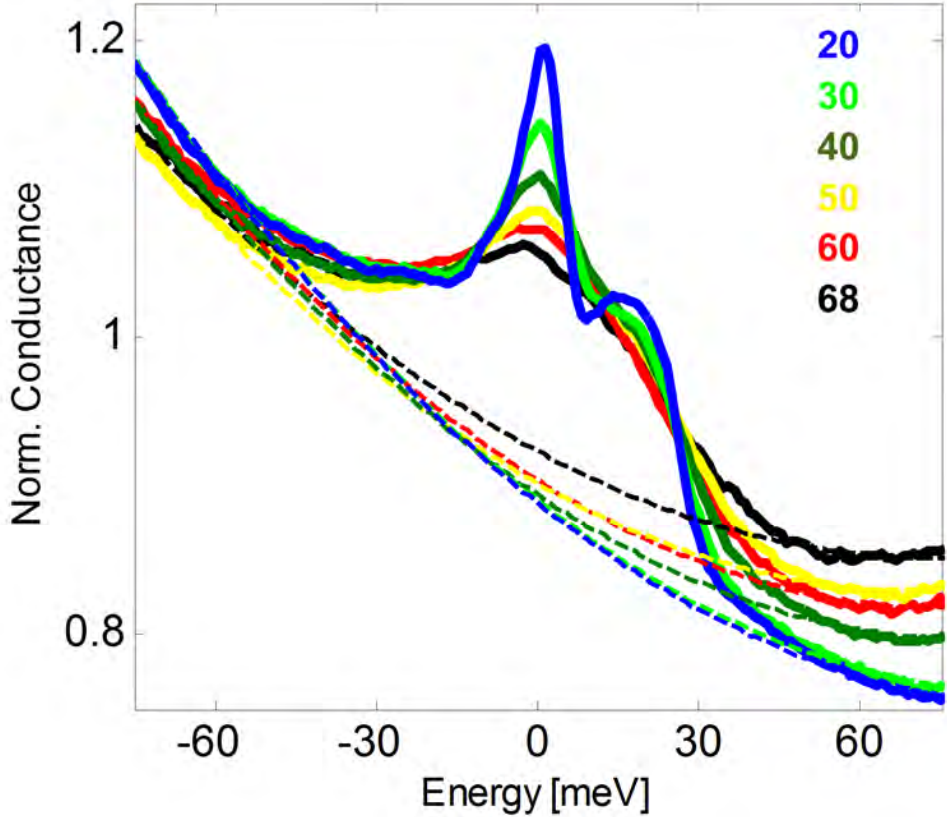

**Fig. S11**: Spectroscopic lineshapes of Fig. 3b (solid lines) together with a second order polynomial fit to the bias range lying outside ±50 meV (dashed lines).

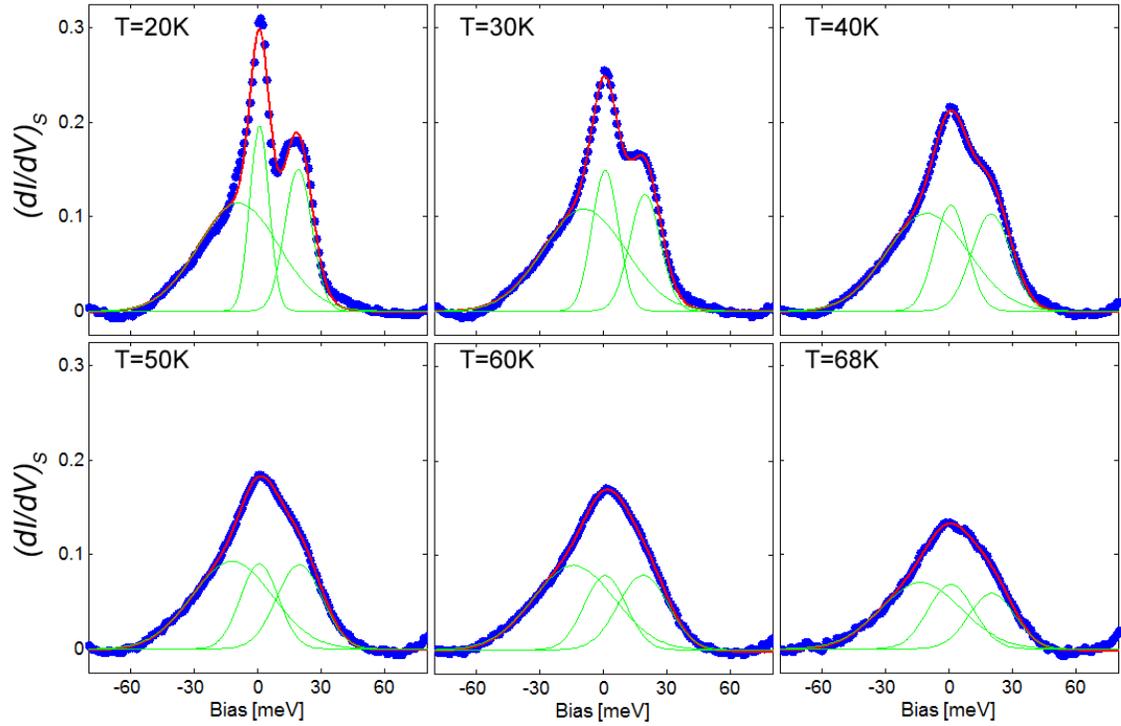

**Fig. S12**: Background subtracted spectra for different temperatures (blue) together with a fit of three Gaussian peaks to the data (red line). The green lines represent the individual Gaussian fit.

## VII. Energy-Temperature Scaling

Figure S13 shows a plot of $(dI/dV)_S * k_B T^\alpha$ as a function of $(E/k_B T)^\beta$ for different values of $\alpha$ and $\beta$. Clearly the plots show that the collapse of the data at different temperatures to a single curve occurs only for $\alpha = 0.53$ and $\beta = 1$. To get an estimate of the "goodness of the collapse" we compute the variance $\chi^2$ of the data at different temperatures (after interpolating the data at each temperature to the same number of data points N=40) with respect to each other as a function of $\alpha$,

$$\chi^2 = \sum_{x=1}^{N} \sum_{T,T'} \frac{[y_T(x) - y_{T'}(x)]^2}{y_{T'}(x)}$$

where $x = (E/k_B T)^\beta$ and $y_T = (dI/dV)_S * k_B T^\alpha$. The insets show the extracted $\chi^2/N$ as a function of $\alpha$ for the three different values of $\beta$.

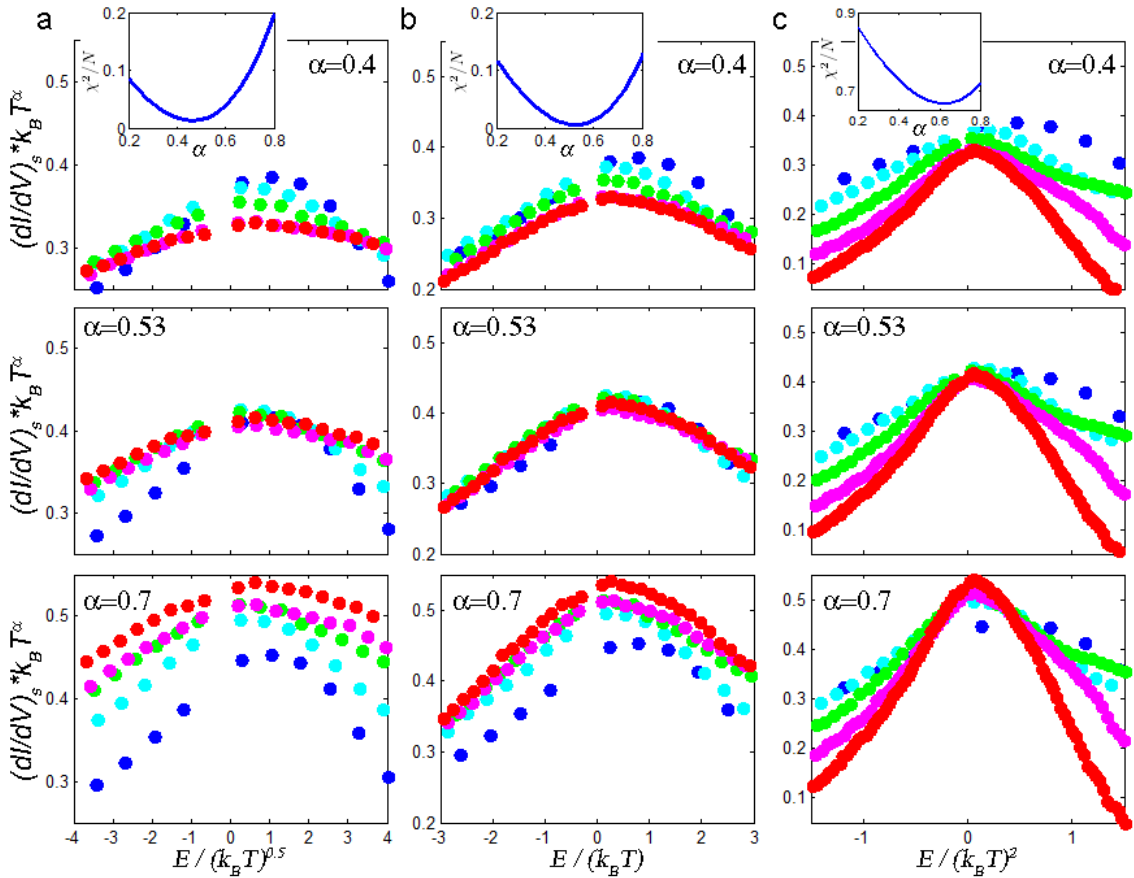

**Fig. S13:** Energy-temperature scaling of the spectra of Fig. 3b of the main text within a narrow energy window near the Fermi energy scaled with **a** $(k_B T)^{0.5}$, **b** $(k_B T)$, and **c** $(k_B T)^2$, for different values of the critical exponent $\alpha$. Insets show the $\chi^2/N$ as a function of $\alpha$ for the three different cases.

In Fig. S14 we plot a map of $\chi^2/N$ as a function of both $\alpha$ and $\beta$. Clearly the map shows that the smallest value of $\chi^2/N$, corresponding to the best collapse of the data, is achieved for $\alpha$ = 0.53±0.03 and $\beta$ = 1.0±0.05.

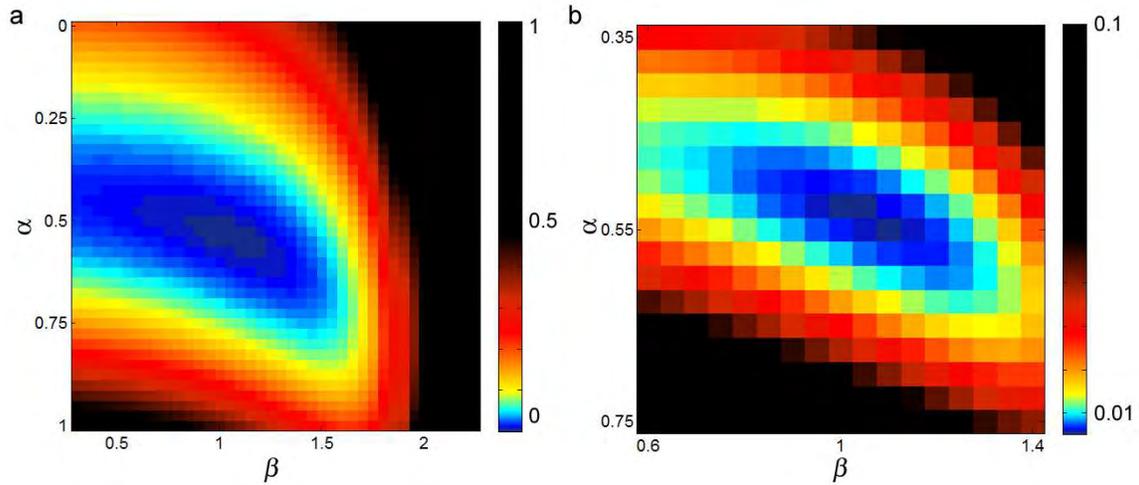

**Fig. S14: a** A map of $\chi^2/N$ as a function of both $\alpha$ and $\beta$. **b** Zoom in of the low $\chi^2/N$ region in **a**.

**References**

**S1**: *Ting Yuan, Jeremy Figgins, and Dirk Morr*, **arXiv:** 1101.2636v1 (2011);
*Jeremy Figgins and Dirk Morr,* Physical Review Letters **104,** 187202 (2010)

**S2**: *P.M. Oppeneer et al.* Journal of Magnetism and Magnetic Materials **310,** 1684 (2007)

**S3:** *P.M. Oppeneer et al.* private communication

**S4:** *Schroder, A. et al.* Nature **407**, 351-355 (2000)

**S5:** *Aronson, M. C. et al.* Physical Review Letters **75**, 725 (1995)